\documentclass[aps,pra,
			   reprint,
			   a4paper,
			   showpacs,
			   superscriptaddress,
			   longbibliography, 
			   floatfix, 
			  ]{revtex4-1}	
\usepackage{datetime} 
\usepackage[colorlinks=true,linkcolor=blue,urlcolor=blue,citecolor=blue]{hyperref}
\usepackage{graphicx}	
\RequirePackage{amsmath}
\RequirePackage{mathtools} 
\usepackage{amsfonts}
\usepackage{amssymb}
\usepackage{amsbsy}
\usepackage{dsfont}   
\usepackage{latexsym}
\def\C60{$\text{C}_{60}$}	
\newcommand{\ket}[1]{\ensuremath{\left|#1\right\rangle}}
\newcommand{\bra}[1]{\ensuremath{\left\langle #1\right|}}
\usepackage{color}	
\begin{document}	
%
\title{Quantum walk transport properties on graphene structures}
\date{\textcolor{red}{\currenttime \,\, \today}}	
%
\author{Hamza Bougroura}
\email{h\_bougroura@hotmail.com}	
\affiliation{Laboratoire de Physique Mathematique et Subatomique (LPMS), University of Constantine~1, Algeria}
\affiliation{Department of physics, Faculty of Exact Science, University of Tebessa, Algeria}
\affiliation{Department of Physics, Durham University, South Road, Durham DH1~3LE, UK}

\author{Habib Aissaoui}
\affiliation{Laboratoire de Physique Mathematique et Subatomique (LPMS), University of Constantine~1, Algeria}
\author{Nicholas Chancellor}
\author{Viv Kendon}
\email{viv.kendon@durham.ac.uk}		
\affiliation{Department of Physics, Durham University, South Road, Durham DH1~3LE, UK}
%
\begin{abstract}
We present numerical studies of quantum walks on \C60 and related graphene structures, to investigate their transport properties.  Also known as a \emph{honeycomb lattice}, the lattice formed by carbon atoms in the graphene phase can be rolled up to form nanotubes of various dimensions.  Graphene nanotubes have many important applications, some of which rely on their unusual electrical conductivity and related properties.  Quantum walks on graphs provide an abstract setting in which to study such transport properties independent of the other chemical and physical properties of a physical substance.  They can thus be used to further the understanding of mechanisms behind such properties.  We find that nanotube structures are significantly more efficient  in transporting a quantum walk than cycles of equivalent size, provided the symmetry of the structure is respected in how they are used. We find faster transport on zig-zag nanotubes compared to armchair nanotubes, which is unexpected given that for the actual materials the armchair nanotube is metallic, while the zig-zag is semiconducting.
\end{abstract}
%
\maketitle

\section{Introduction\label{sec:intro}}
Quantum versions of random walks have been extensively studied for the past three decades, leading to a range of applications.  In this paper we focus on their transport properties.  Their potentially exponential quantum speed up over equivalent classical random walk transport was first proved in an algorithmic setting by \citeauthor{kempe02a} \cite{kempe02a,kempe02c} on a hypercubic graph, followed by \citeauthor{childs02a} \cite{childs02a} on a specially chosen ``glued trees'' graph.  In a physical setting, quantum transport on spin chains \cite{bose03a} is isomorphic to continuous-time quantum walks.  This lead to intensive study of how to optimise quantum state transfer over short chains, as reviewed by \citeauthor{kay09a} \cite{kay09a,kay10a}, with applications as quantum wires to connect components in quantum devices for communication and computation.  Quantum walks can also reproduce the phenomenon of Anderson localisation \cite{anderson58a,keating06a,krovi05a}, highlighting the importance of controlling the quantum walk parameters to achieve efficient transport.  Studies by \citeauthor{krovi06a} \cite{krovi06a,krovi07a} expose the role of symmetry in the underlying graph structures in quantum walk transport.  

Noting the importance of graphene and related substances as materials with many interesting electrical properties \cite{Novoselov04a,Novoselov05a,White98a,Mintmire92a,Hamada92a,Saito92a} in this paper we apply quantum versions of random walks to study quantum transport properties on various structures based on \C60 and graphene lattices.  The variation in conductivity of graphene is exploited in diverse application \cite{Dale16a,Tans98a,Bachtold01a,Woo09a,Clegg14a}.  Recently, continuous-time quantum walks on graphene lattices have been studied by \citeauthor{foulger15a} \cite{foulger15a} to implement a quantum walk search algorithm, and apply this to communications between selected nodes on the lattice.  The lattice is also called ``honeycomb'' and other studies of quantum walks on this type of graph can be found in \cite{Lyu15a,Jafarizadeh07a,abal10a}.

Analytical solutions for quantum walks on graph structures are challenging.  Even the simplest cases of the cycle \cite{aharonov00a}, hypercube \cite{moore01a} and Cartesian lattices \cite{ambainis01a,grimmett03a,gottlieb04a} require significant mathematical effort.  For perfect state transfer on small graphs, where the quantum state of the walker is exactly reproduced at the target node, analytical results are known for a few special cases, reviewed by \citeauthor{kendon10b} \cite{kendon10b}.  This is also because perfect state transfer is hard to achieve in general, and for many applications it is sufficient to obtain fast probabilistic transfer.  Analytical solutions usually require a homogeneous graph structure, either finite or infinite, or parameter tweaking for each situation \cite{kay10a}.  There are techniques to compose compatible small graph structures into larger ones \cite{canul09a} for which the analytical solutions can also be combined, but there are limitations to this method.  For studies on more general structures, numerical simulation is the best option, allowing the range of structures to be extended to be more practically relevant.  It is also generally the case that discrete-time and continuous-time quantum walks give similar results on the same graphs, and their equivalence has been shown analytically for the line \cite{strauch06a} using a method that can be expected to generalise for other homogeneous lattices.  While the continuous-time walk can be more tractable analytically, for numerical simulation the unitary operators of the discrete-time walk are more convenient than the numerical integration required for the continuous-time walk.  We therefore carried out our studies using the discrete-time quantum walk.  Discrete, coined quantum walks can be implemented using atoms trapped in optical lattices, for example, and honeycomb lattices can be created this way.  Internal degrees of freedom of the atoms then play the role of the quantum coin, by coupling to the direction of motion of the atoms as the optical lattice is modulated \cite{satapathy09a,cote06a,tarruell12a,polini13a}.

The paper is organised as follows.  In section \ref{sec:qw} we define our model of a quantum walk and discuss its behaviour on cycles.  In section \ref{sec:results} we apply the quantum walks to \C60 and graphene nanotubes, focusing on the efficiency of transport between specified points on the structures.  In section \ref{sec:conc} we summarise our findings and discuss their applications and directions for future research.

\section{Quantum walks on graphs\label{sec:qw}}
We define a discrete-time, coined, quantum walk on a regular, connected, undirected graph $G$ as follows.  First we specify the graph on which the quantum walk takes place.  For a graph $G$ with $n$ nodes, let $V$ be the set of nodes and $E$ be the set of edges connecting pairs of nodes.  We label each node with a unique number $v\in\mathbb{Z}_n$, and identify the edges by the labels of the nodes they connect.  Thus, for $u,v\in V$, we have $(u,v)\in E$ iff there is an edge $(u,v)$ connecting node $u$ to node $v$.  Graph $G$ is undirected, i.e., $(u,v) \equiv (v,u)$ and connected, i.e., $\forall u,v \in V$ $\exists \{w\}$ $| (w_i,w_{i+1}) \in E$ $\forall i \in \{1... |\{w\}|-1\}$  
, and there is at most one edge between any pair of nodes.  The degree $d(v)$ of node $v\in V$ is the number of edges meeting at $v$.  For a regular graph of degree $d$ we have $d(v) = d$ $\forall v\in V$.  In order to support the dynamics of the quantum walk, at each node $v\in V$ we label the ends of the edges at that node from $0\dots (d-1)$ in an arbitrary but fixed order.  For $a,b\in\{0\dots (d-1)\}$, we define an edge label function
\begin{equation}
	e(u,a)=(v,b)
	\label{eq:edge_label_function}
\end{equation}
that returns the ordered pair of labels at the other end of the $a$th edge at node $u$, i.e., the $b$th edge at node $v$.

A discrete-time, coined quantum walk on a regular, connected, undirected graph $G$ has a discrete Hilbert space $H=H_G\times H_d$ where $H_G$ has dimension $n$ corresponding to the number of nodes, and $H_d$ has dimension $d$, corresponding to the number of edges $d$ meeting at each node.  We choose a natural and convenient set of basis states $\ket{j,c}\equiv\ket{j}\otimes\ket{c}$ with $j\in\mathbb{Z}_n$ and $c\in\mathbb{Z}_d$.  A quantum walker can thus be thought of as a particle with an internal degree of freedom of dimension $d$ that is located on a node, or in superposition on nodes, of the graph.  A general state $\ket{\psi(t)}$ of the quantum walker at time $t$ can be written
\begin{equation}
	\ket{\psi(t)} = \sum_{j\in V, c\in \mathbb{Z}_d}\alpha_{j,c}(t)\ket{j,c},
\end{equation}
where the coefficients $\alpha_{j,c}(t)$ are complex amplitudes normalised such that $\sum_{j,c}|\alpha_{j,c}(t)|^2=1$.
The dynamics of the quantum walk are unitary in discrete unit time intervals 
\cite{*
[{A continuous Hamiltonian evolution that reproduces the discrete, unitary dynamics can be obtained, see for example, }] [{.  While relevant for experimental implementations, it is not useful for this study.}] kendon04a}.
We 
utilize a mixture of \textbf{no-flip-flop} steps and \textbf{flip-flop} steps to build the 
shift operator elements of each structure, except in the case of the 
cycle, where we utilize purely the no-flip-flop steps to build the shift 
operator $S$. For the benefit of those readers who are unfamiliar with the action of the \textbf{flip-flop} shift operator, we will explain its action in detail for clarity. The operator $S$ acts on both the coin state and position of the walker to move it between nodes that are connected by edges.  It is defined by its action on the basis states,
\begin{equation}
	S\ket{u,a} = \ket{e(u,a)} = \ket{v,b},
\end{equation}
making use of the edge label function defined in eq.~(\ref{eq:edge_label_function}).   We note that $e(e(u,a))=(u,a)$, hence $S.S\ket{u,a}=\ket{u,a}$, confirming that $S$ is its own inverse, and therefore unitary.
A coin operator $C$ acts only on the coin degrees of freedom.  We are free to choose $C$ to be any unitary operation of dimension $d$.  There are some natural choices for $C$ that we will introduce and discuss later.  The role of the coin operator is equivalent to tossing the coin in a classical random walk: it rearranges the amplitudes for different coin states.
A single step of the quantum walk consists of a coin operation followed by a shift, giving
\begin{equation}
	\ket{\psi(t+1)} = S.\left(\openone_n\otimes C\right)\ket{\psi(t)},
\end{equation}
where $\openone_n$ is the identity operation on the position space $H_n$ of the quantum walker.  A quantum walk of $T$ steps from an initial state of $\ket{\psi(0)}$ can be written
\begin{equation}
	\ket{\psi(T)} = \biglb(S.\left(\openone_n\otimes C\right)\bigrb)^T\ket{\psi(0)}.
\end{equation}

The initial state has a significant impact on the subsequent quantum walk, unlike for a classical random walk, where the initial state is irrelevant to the long time behavior.  This is because the quantum walk is a deterministic, unitary dynamics.  The range of choices for the initial state is large, and we are interested in transport properties that are not particularly sensitive to the choice of initial state.  For the quantum walk on the line, for example, the spreading rate is linear, regardless of the initial state \cite{ambainis01a}.  For the studies presented here, we used unbiased, symmetric initial states, either in terms of the coin states at a single node, or an equal superposition of such states on a group of neighboring nodes.

\begin{figure}
	\includegraphics[width=1.0\columnwidth]{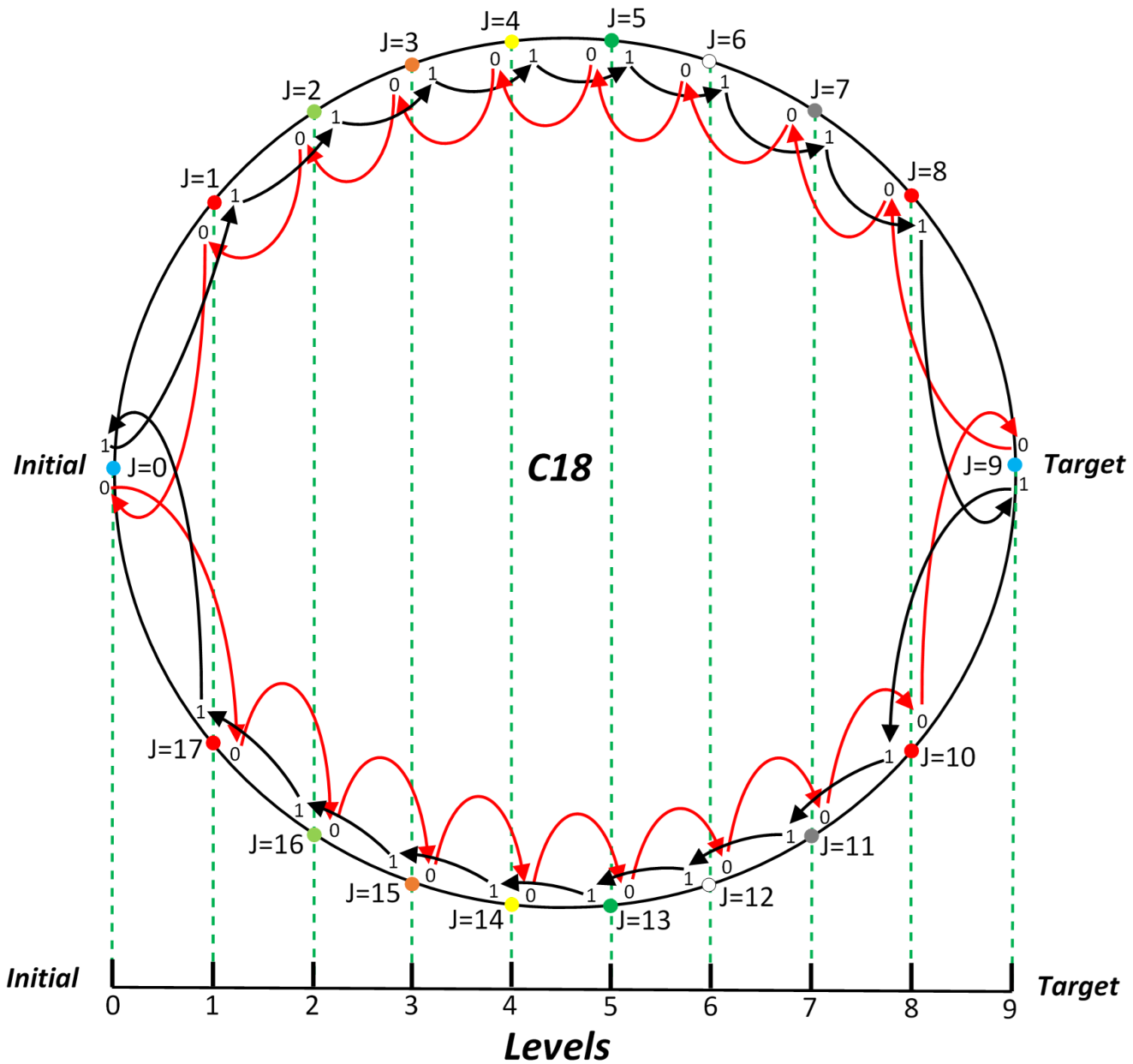}
	\caption{A cycle $C_{18}$ with 18 nodes, showing a convenient set of node (j) and edge ($\{0,1\}$) labels as described in the text. The black and the red arrows represent  respectively the two possible steps ($\ket{j+1,1}\bra{j,1}$) and ($\ket{j-1,0}\bra{j,0}$) for each node (j) in the shift operator $S_C$.The two opposite nodes are designated `Initial' and `Target', and the nodes between them are mapped in pairs (as colored, shaded in grey in print) to a set of levels, so the progress towards the target node can be quantified.\label{fig:cycle}}
\end{figure}

To motivate our choices of coin operators, we first consider a quantum walk on one of the simplest small graph structures, the cycle.  A cycle $C_n$ with $n$ nodes has a set of edges $\{(j,j+1)\}$ with $j\in\mathbb{Z}_n$ and addition modulo-$n$, so that node $(n-1)$ is connected to node $0$.  The shift operator is thus
\begin{equation}
S_C = \sum_{j}\biglb(\ket{j+1,1}\bra{j,1} + \ket{j-1,0}\bra{j,0}\bigrb),
\end{equation}
where we have used a consistent labeling of the nodes and ends of the edges such that the label $1$ is on the end of the edge that connects node $j$ to node $j+1$ and vice versa for label $0$, see figure \ref{fig:cycle}.  This choice of labels is not necessary, but it does simplify the analysis, both numerical and analytical.
The cycle has degree $d=2$, so we need a 2-dimensional coin operator.  Ideally, we also want our coin operator to be unbiased, so that no matter which direction the walker arrived from it has an equal chance of leaving by either edge.  We can achieve this by using the Hadamard operator
\begin{equation}
	H = \frac{1}{\sqrt{2}}\bigglb(\begin{array}{rr}1&1\\1&-1\\\end{array}\biggrb).
	\label{eq:hadamard}
\end{equation}
A symmetric form
\begin{equation}
	H_{i} = \frac{1}{\sqrt{2}}\bigglb(\begin{array}{rr}1&i\\i&1\\\end{array}\biggrb)
\end{equation}
can also be used.  In general, a phase factor of $\pi$ ($-1$ and $i\times i$ in the above, respectively) can be distributed in various equivalent ways, with corresponding cosmetic changes in the quantum walk \cite{bednarska03a}.

The discrete-time quantum walk on the cycle was solved analytically by \citeauthor{aharonov00a} \cite{aharonov00a}, and its properties are well-studied.  Of particular note for our purposes is its use in quantum state transfer.  Using cycles of even-$n$ size, the quantum walk starts at one node with the aim of reaching the opposite node.  First noted for $n=4$ by \citeauthor{travaglione01a} \cite{travaglione01a}, this can provide perfect state transfer for suitably chosen coin operators (not always the unbiased Hadamards) \cite{tregenna03a,dukes14a} and initial states.  We used a corresponding size of $n$-cycle to provide a benchmark for evaluating the performance of quantum walk transport on graphene structures.

\subsection{Numerical methods\label{ssec:numethods}}
Our numerical simulation code was written in Python 3.5 using the NumPy, SciPy, and Matplotlib packages \cite{python,scipy,numpy,matplotlib}.  Most of the simulations took no more than a few minutes each on standard desktop computers, so no special numerical optimisation techniques were required.  Figures of the \C60 and nanotube structures were drawn using Virtual NanoLab \cite{vnl}.

\subsection{Transport measures\label{ssec:transmeas}}
There are many possible properties of quantum walks that can be calculated.  For the smallest simulations, we visualised the probability distribution step-by-step to obtain a detailed picture of the behaviour.  We then calculated the average position over time, $\langle x \rangle$, where $x$ is the position mapped to levels as shown in figure \ref{fig:cycle},
\begin{equation}
\langle x \rangle = \sum_{j,c} x(j) |\alpha_{j,c}(t)|^2
\end{equation}
We also calculated the accumulated arrival probability $A(T)$.  The accumulated arrival probability is equivalent to putting a ``sink'' at the target node, and summing the probability of the walker being in the sink after each step of the quantum walk
\begin{equation}
	A(T) = \sum_{t=0}^{T} \sum_{c}|\alpha_{a,c}(t)|^2
\end{equation}
where $a$ is the target node, and $\alpha_{c,a}(t)$ is reset to zero before the next step of the quantum walk is applied.  This is a non-unitary process with a practical operational interpretation.  After each step of the quantum walk, the target node is measured to check for the presence of the walker.  It will be found with probability $|\alpha_{c,a}(t)|^2$.  With probability $1-|\alpha_{c,a}(t)|^2$ the walker is somewhere else on the graph, and the quantum walk continues to evolve, but without the amplitude on node $a$, because we just found out it isn't there.  Note that $A(T)$ is a monotonically-increasing function of time, because once the quantum walker has arrived at node $a$, it doesn't leave it.  After comparing $\langle x \rangle$ with $A(T)$, we chose the latter as the clearest indicator of successful quantum transport.  

\begin{figure}
	\includegraphics[width=1.0\columnwidth]{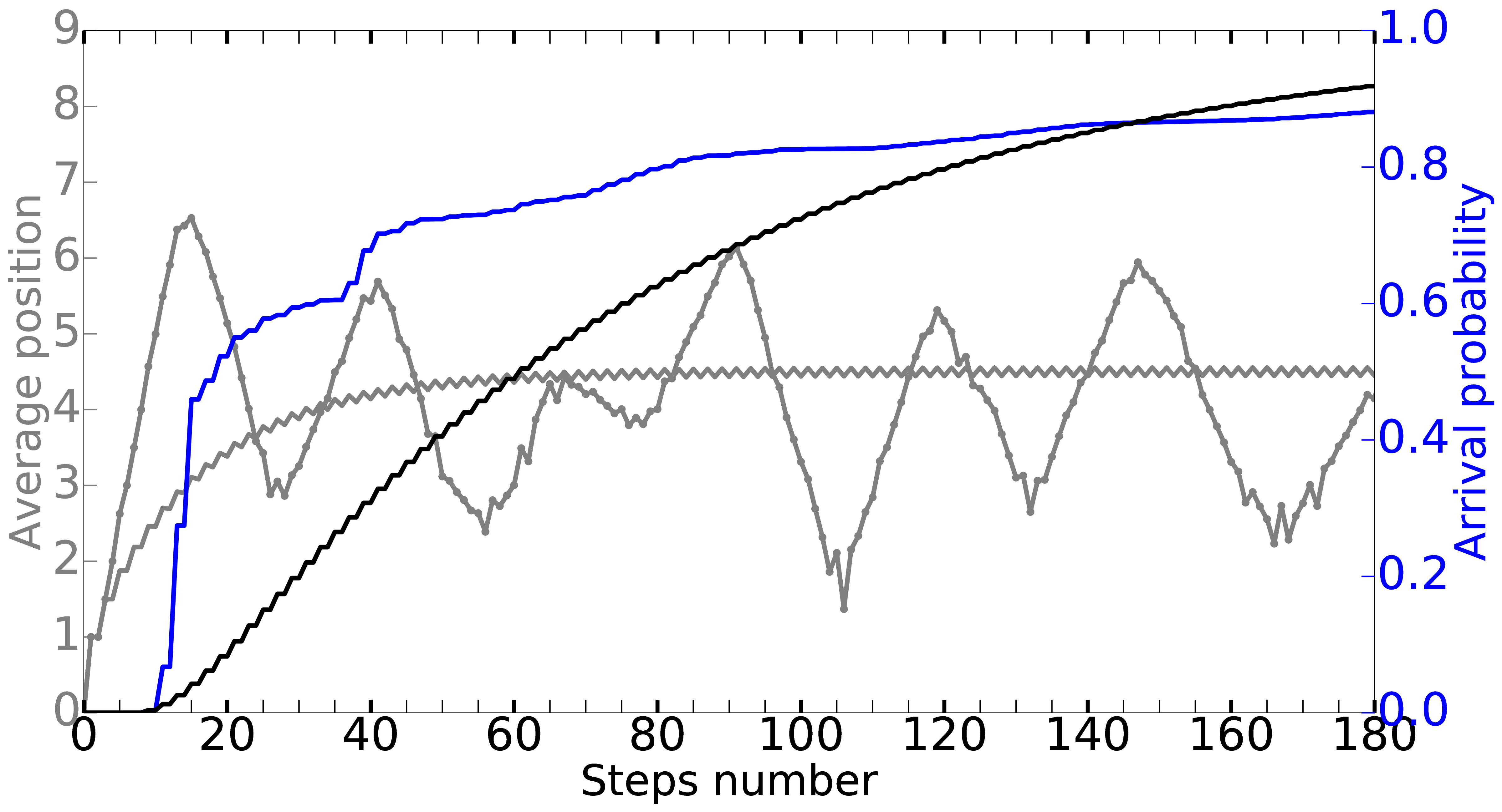}
	\caption{(Color online.) 180 steps of a quantum walk on a cycle $C_{18}$ with $18$ nodes mapped to ten levels from start ($0$) to target node ($9$), see figure \ref{fig:cycle}.  Average position (grey, left axis) and arrival probability (blue, right axis, dark grey in print) plotted against the number of time steps for coin operator $H$ (solid). A classical random walk using an unbiased 2-sided coin (black) shown for comparison.\label{fig:qwcycle}}
\end{figure}

With the sink at the target site, analytical solution is even more challenging.  For the cycle, a numerical comparison of average position and arrival probability is shown in figure \ref{fig:qwcycle}, along with the equivalent quantities for a classical random walk, for comparison.
For short times, the quantum walk arrives sooner, and the arrival probability approaches unity faster, than the classical random walk.  For longer times, the curves cross and the classical random walk approaches unity faster.  For very long times, both asymptote to unity (see supplementary material).  In general, a quadratic speed up is expected for quantum walks on the cycle when compared with a classical random walk \cite{magniez09a}.  This is an asymptotic result for large cycles and does not apply directly to small cycles like $C_{18}$.  The short time behavior, a steep rise in the arrival probability in the first 20 time steps, is the quantum speed up in this instance.

\subsection{Coin operators\label{ssec:coins}}
\begin{figure}
	\includegraphics[width=1.0\columnwidth]{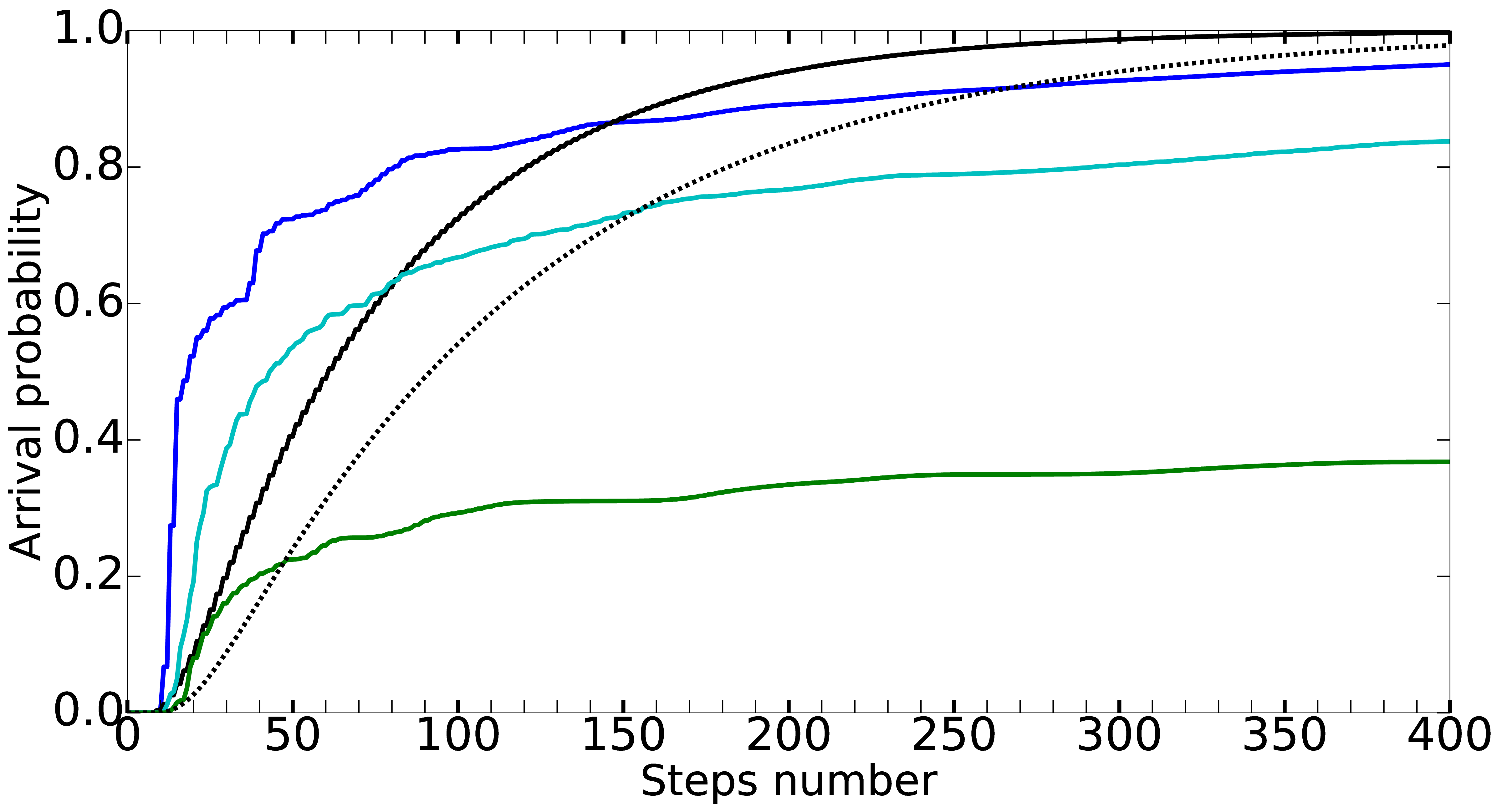}
	\caption{(Color online.) A quantum walk of 400 steps on a cycle $C_{18}$ with $18$ nodes mapped to ten levels from start ($0$) to target node ($9$), see figure \ref{fig:cycle}.  Arrival probability plotted against the number of time steps for coin operators $H$ (blue, dark grey in print), $G_3$ (green, mid grey in print), $F_3$ (turquoise, light grey in print).  A classical random walk using an unbiased 2-sided (black, solid) and 3-sided coin (black, dotted) shown for comparison.\label{fig:qwcycle3d}}
\end{figure}

There is another natural choice for the coin operator that models many realistic situations.  As well as shifting to a connected node at each time step, the quantum walker may have a third choice, to stay at the current node.  This can be achieved using a coin of dimension $c=d+1$.  Thus, for the cycle, this needs a coin of dimension three.  We tested the walk on the cycle using a coin operator known as a Grover coin operator.  The Grover coin operator can be defined for any dimension $d\ge 2$,
\begin{equation}
G_d = \frac{1}{d}\left(\begin{array}{cccc}2-d&2&\cdots&2\\2&2-d&\cdots&2\\\vdots&\vdots&\ddots&\vdots\\2&2&\cdots&2-d\\\end{array}\right).
\end{equation}
For $d=2$ it reduces to the Pauli $\sigma_x$ operator, which corresponds to steps in a single direction (completely biased).  In three dimensions, it is
\begin{equation}
G_3 = \frac{1}{3}\Bigglb(\begin{array}{rrr}-1&2&2\\2&-1&2\\2&2&-1\\\end{array}\Biggrb).
\label{eq:g3coin}
\end{equation}
The Grover coin operator has the highest degree of symmetry possible in a unitary operator.  The incoming direction is already special, but all other directions are treated exactly the same for both amplitude and phase of the outgoing state.  The arrival probability for the quantum walk with a $G_3$ coin operator is shown in figure \ref{fig:qwcycle3d}. 

Another three-dimensional coin operator is the Fourier coin operator, which can also be defined for any dimension,
\begin{equation}
F_d = \frac{1}{\sqrt{d}}\left(\begin{array}{cccc}1&1&\cdots&1\\1&\omega&\cdots&\omega^d\\\vdots&\vdots&\ddots&\vdots\\1&\omega^d&\cdots&\omega^{d\times d}\\\end{array}\right),
\end{equation}
where $\omega=e^{2\pi i/d}$ is the complex $d$th root of unity.  For $d=2$, the Fourier coin operator reduces to the Hadamard coin operator in equation (\ref{eq:hadamard}).  The Fourier coin operator is unbiased for all dimensions, i.e., the walker is equally likely to leave by any available edge, regardless of which it arrived along.  However, this comes at a cost of the relative phases being different for each direction, to ensure the coin operator is unitary overall.  Since the phase factors are what gives the quantum walk its advantage over the classical random walk, the Fourier coin operator can produce very different quantum walk behaviors when compared with the Grover coin operator.  Using $F_3$ for a cycle with a ``wait state'' is also shown in figure \ref{fig:qwcycle3d}.  Out of all these choices for coin operator, we can see the Hadamard coin is the fastest, both $H$ and $F_3$ beat the classical random walk for short times, while $G_3$ only beats the classical three-sided coin for short times, and in fact never reaches the opposite side with certainty, see appendix.

\section{Results\label{sec:results}}
We now present our results for the transport properties of quantum walks on \C60 and various carbon nanotube structures.  First, the different possible coin operators are compared on \C60.  Then the role of symmetry in the structures themselves is investigated, by using different combinations of starting nodes and target nodes.  Finally, we studied nanotubes made from cylinders of graphene, both with capped ends and as loops, to explore how the width and length affect the overall transport efficiency.

\subsection{Comparison of coin operators on \protect{\C60}\label{ssec:coinops}}
\begin{figure}
	\begin{minipage}{\columnwidth}
		\includegraphics[width=0.7\columnwidth]{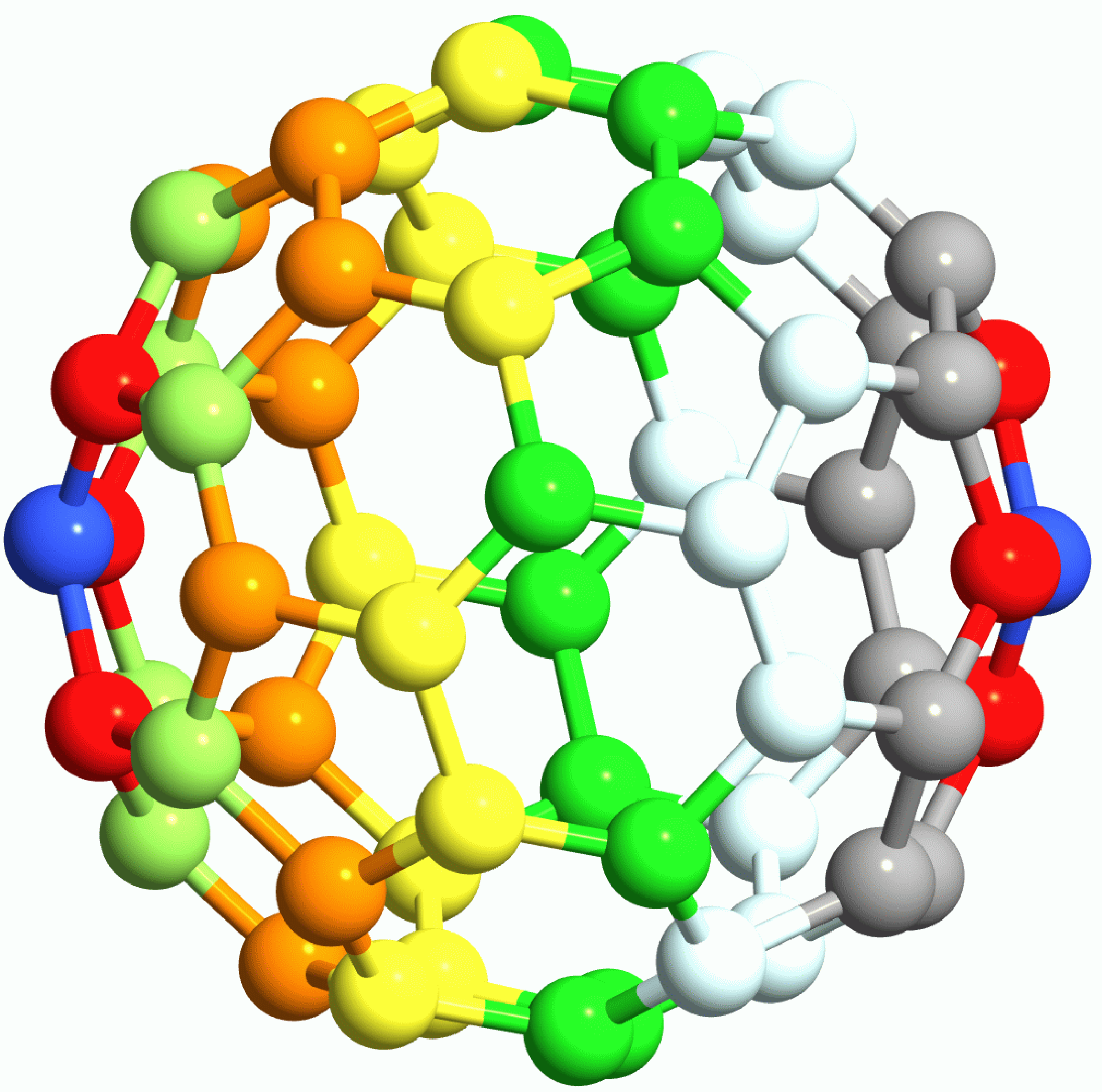}
	\end{minipage}\\
	\caption{(Color online.)  A \protect{\C60} colored to indicate the levels from the initial node (blue, left, darkest grey in print): level 1 (red, left, second darkest grey in print), level 2 (lime, fifth lightest grey in print), level 3 (orange, fourth darkest grey in print), level 4 (yellow, lightest grey in print), level 5 (green, sixth darkest grey in print), level 6 (white), level 7 (grey, third darkest grey in print), level 8 (red, right, second darkest grey in print), level 9 (target, blue, darkest grey in print). \label{fig:c60levels1}}
\end{figure}

\begin{figure}
	\includegraphics[width=1.0\columnwidth]{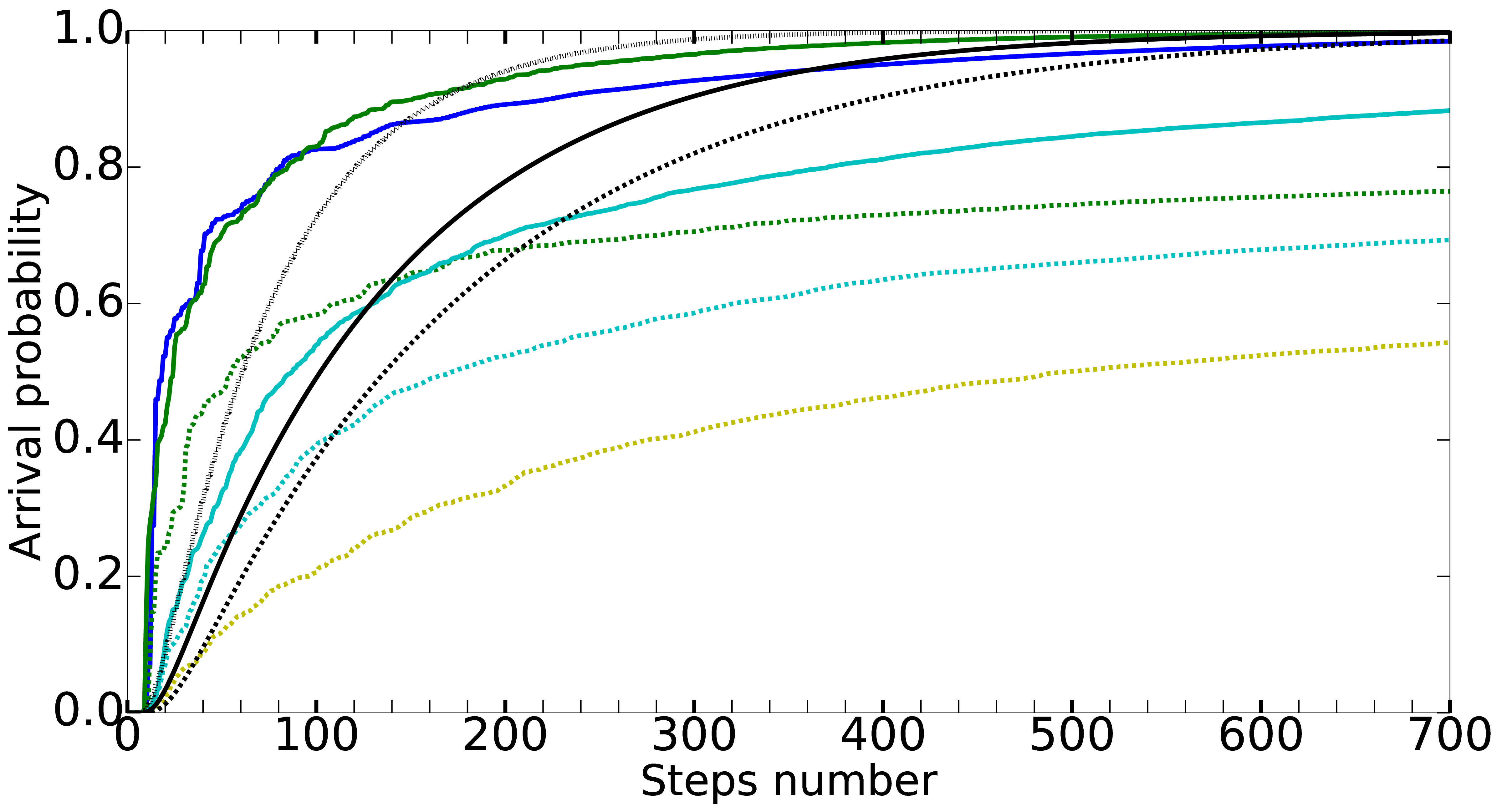}
	\caption{(Color online.) A quantum walk of 700 steps on \protect{\C60} between single opposite nodes, mapped to ten positions as in figure \ref{fig:c60levels1}.  Arrival probability plotted against the number of time steps for coin operators $G_3$ (green, solid, mid grey in print), $G_4$ (green, dashed, mid grey in print), $F_3$ (turquoise, solid, light grey in print), $F_4$ (turquoise, dashed, light grey in print), $H\otimes H$ (yellow, dashed, very light grey in print).  A cycle $C_{18}$ is shown (blue, dark grey in print) along with classical random walks for a 3-sided (black, solid) and 4-sided (black, dashed) coin and the cycle (black, dotted) are shown for comparison.\label{fig:qwC60allcoins}}
\end{figure}

The graphene structures that are the focus of our work all have nodes with $d=3$.  Our first investigation was to compare several natural choices of coin operators, to see how sensitive the transport properties are to different coin operations.  Clearly, the $G_3$ coin of equation (\ref{eq:g3coin}) is an obvious choice, along with $F_3$.  We can also choose a $c=d+1=4$ coin with a ``wait state'', with corresponding four-dimensional coin operators.  We tested $G_4$ and $F_4$ and also a tensor product of two Hadamard coin operators, $H\otimes H$.  In a similar manor as for the cycle, we chose an initial node on the \C60 structure and designated the opposite node the target node, see figure \ref{fig:c60levels1}.  The intermediate nodes are mapped to levels corresponding to the number of edges traversed on the shortest path from the initial node. 
Figure \ref{fig:qwC60allcoins} shows a quantum walk on this \C60 structure from one node to the opposite node.
From the graph, we can see that the $G_3$ coin operator was consistently the best choice for transport properties, outperforming all the other walks even for long times.  We have therefore focused on this coin operator for presenting the results in the following sections.  We tested other coin operators on all structures, and found their performance to be essentially the same in relation to $G_3$ as in figure \ref{fig:qwC60allcoins}.  We also tested for more time steps, see appendix, to confirm that all the $d=3$ coins and the classical random walks do eventually arrive with unit probability, while the quantum walks with wait states do not.

\subsection{Role of symmetry in transport on \protect{\C60}}
\begin{figure}
	\begin{minipage}{\columnwidth}
		(a)
		\includegraphics[width=0.7\columnwidth]{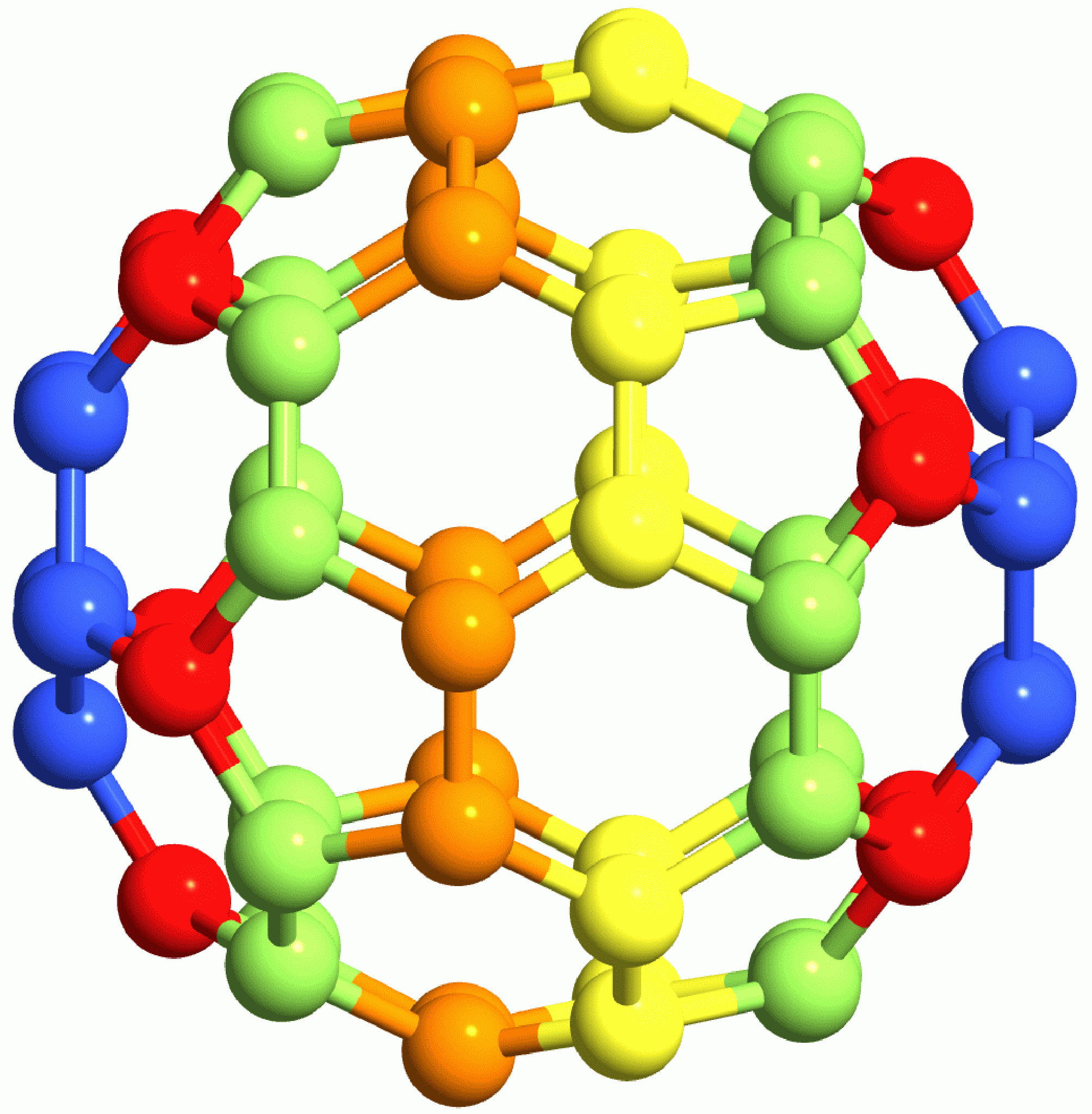}
	\end{minipage}\\
	\begin{minipage}{\columnwidth}
		(b)
		\includegraphics[width=0.7\columnwidth]{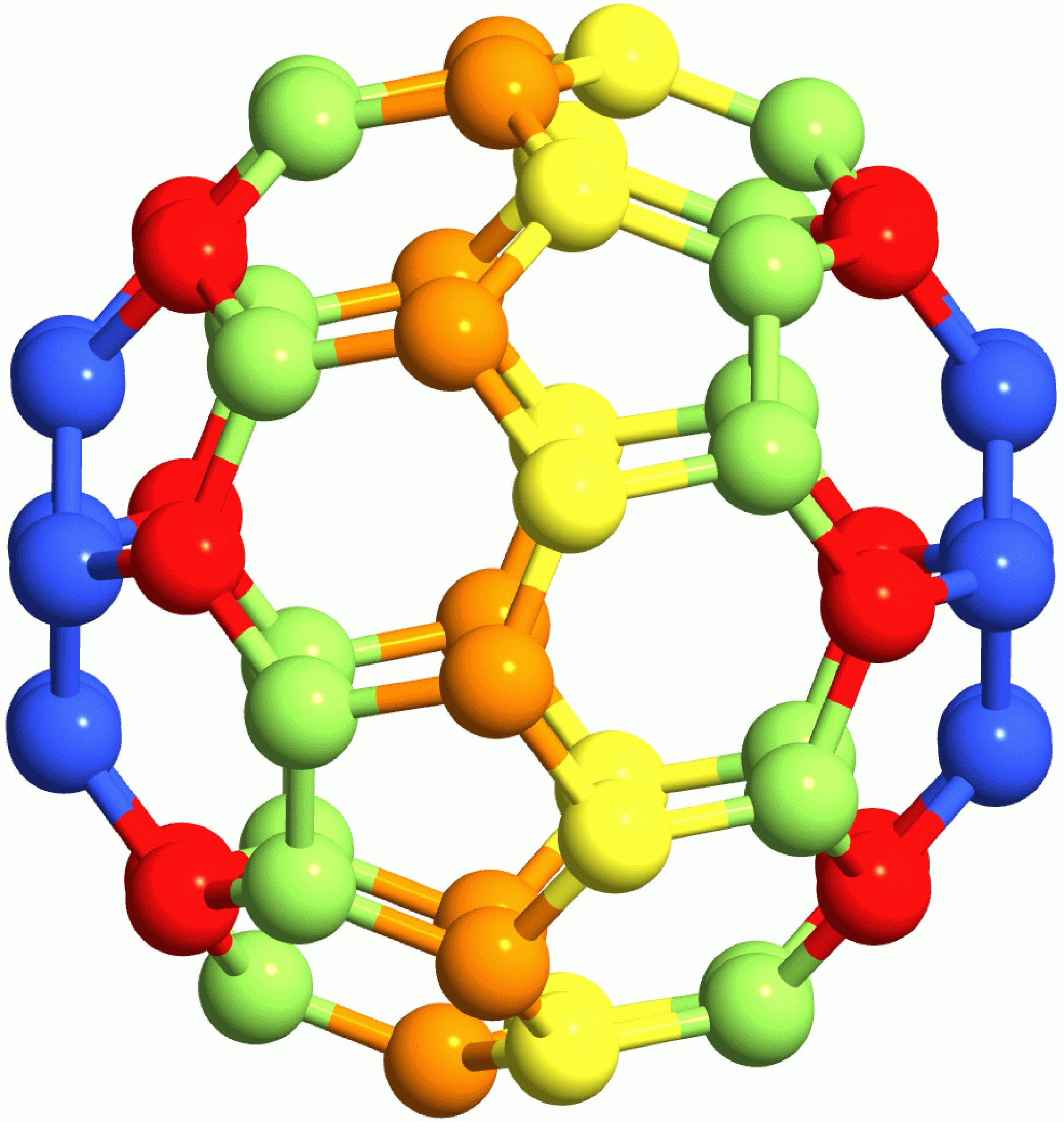}
	\end{minipage}\\
	\caption{(Color online.) \protect{\C60} with different initial positions (a) an equal superposition of all nodes of a pentagonal face, and (b) an equal superposition of all nodes of a hexagonal face.  The nodes between the initial and target faces are colored (shaded in grey in print) to indicate their grouping into eight levels.\label{fig:C60}}
\end{figure}

\begin{figure}
	\includegraphics[width=1.0\columnwidth]{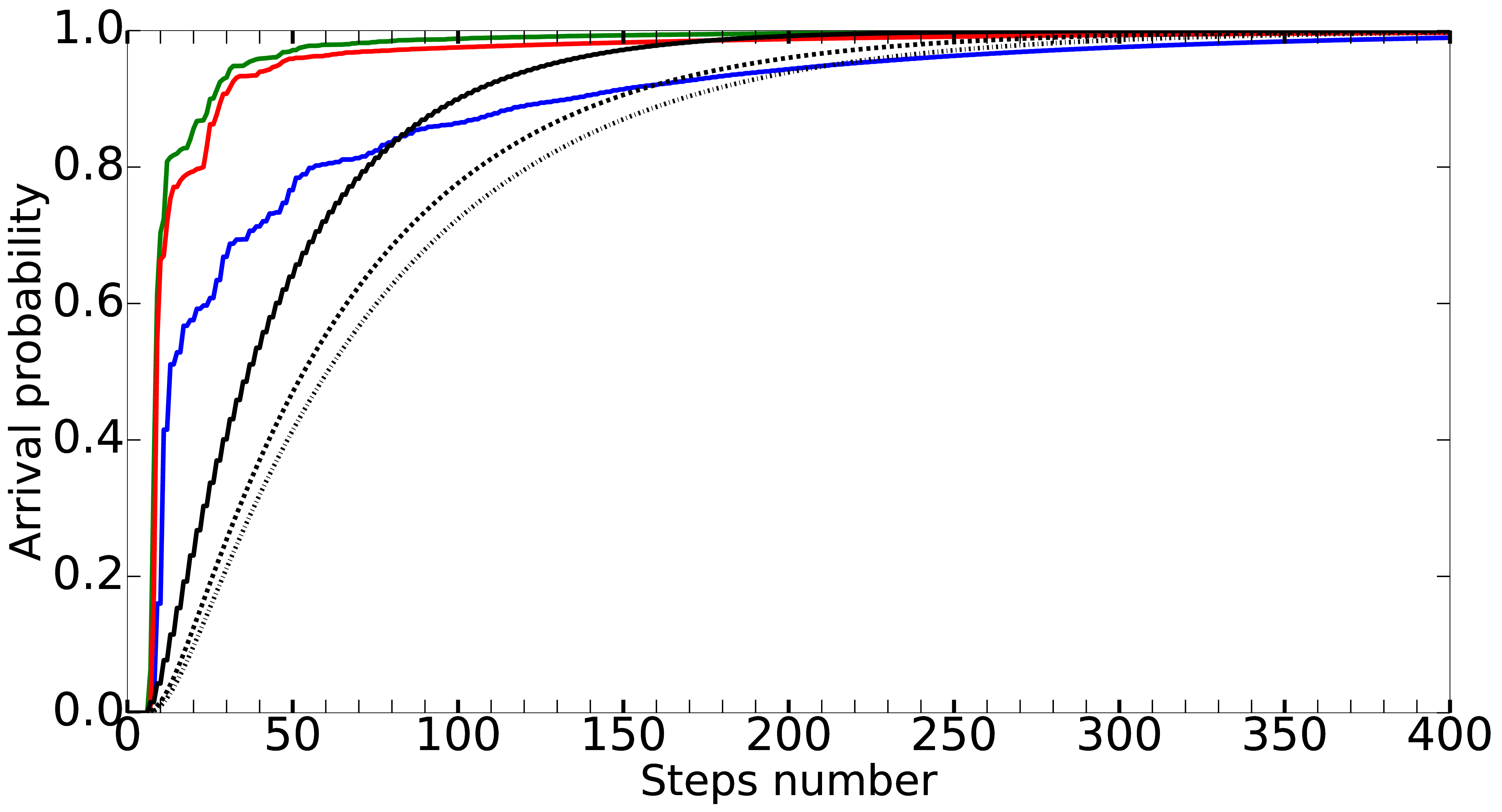}
	\caption{(Color online) A quantum walk of 400 steps on \protect{\C60} mapped to 8 levels using the $G_3$ coin comparing different starting positions: hexagon (green, upper, darker mid grey in print), pentagon (red, lower, lighter mid grey in print). A quantum walk with the $H$ coin on a cycle $C_{14}$ (blue, dark grey) is also shown.  The corresponding classical random walks (black dashed, dotted, solid) are shown for comparison.\label{fig:qwC60}}
\end{figure}

\begin{figure}
	\begin{minipage}{0.48\columnwidth}
		\includegraphics[width=0.9\columnwidth]{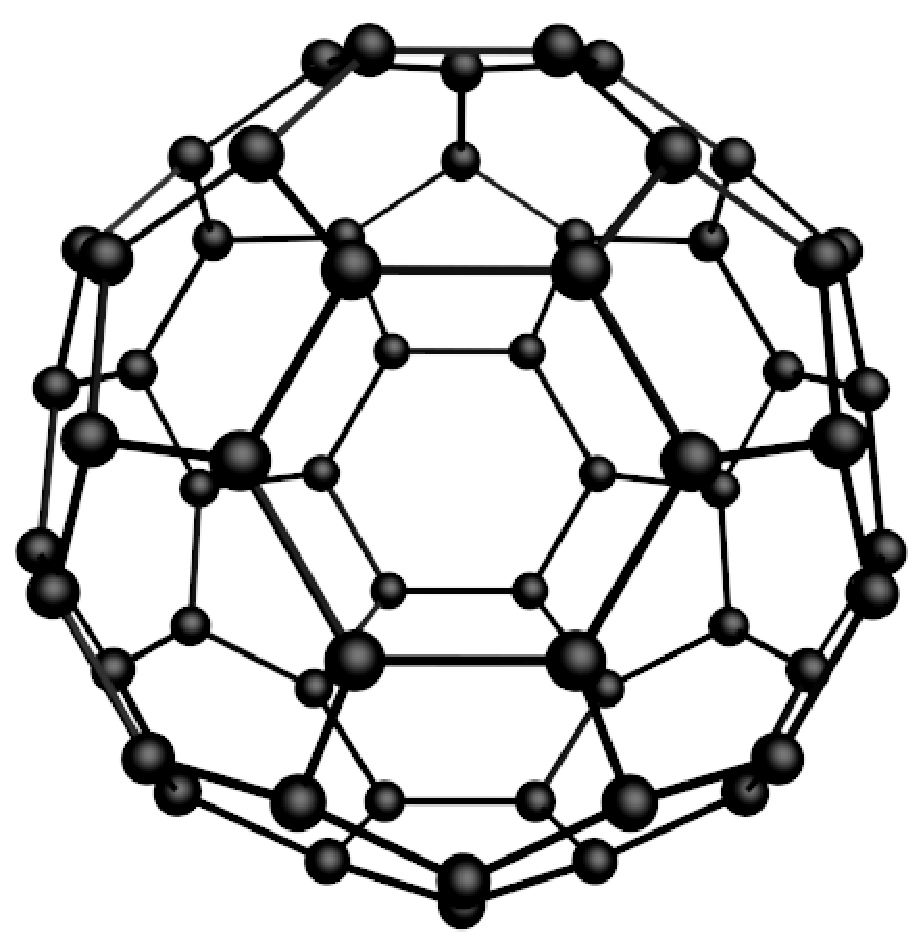}
	\end{minipage}
	\hfill
	\begin{minipage}{0.48\columnwidth}
		\includegraphics[width=0.9\columnwidth]{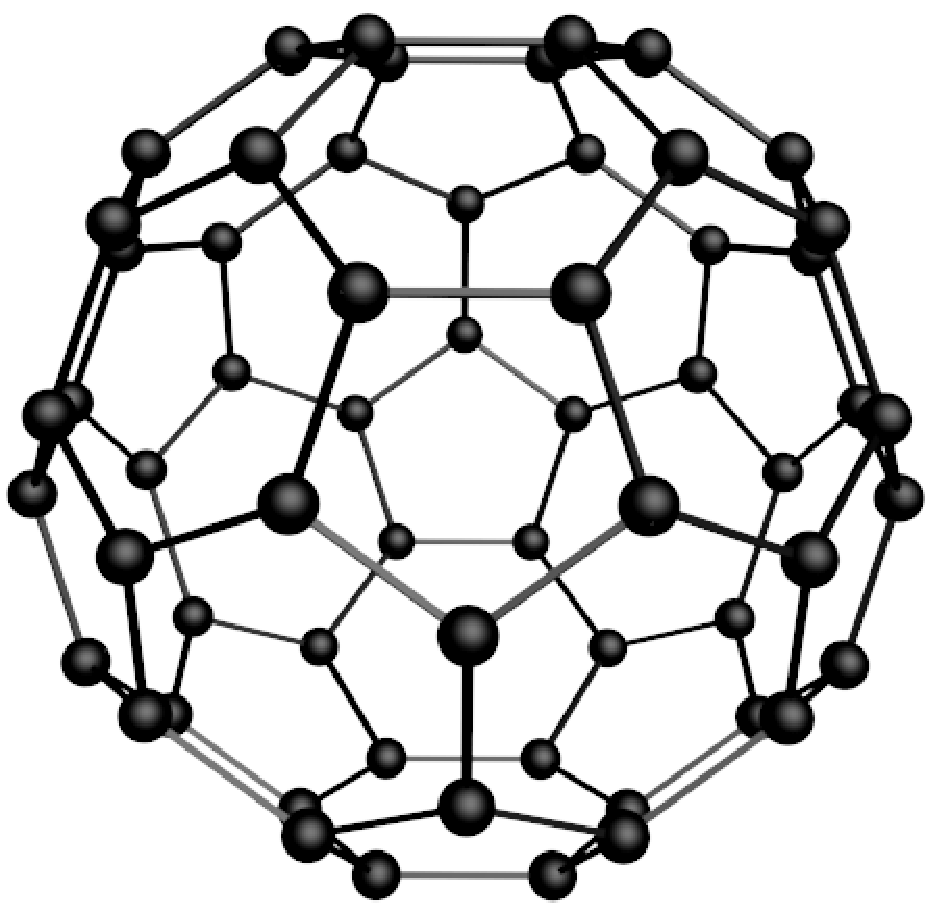}
	\end{minipage}
	\caption{\protect{\C60} viewed end on to illustrate how the hexagonal initial/target pair (left) are more symmetric than the pentagonal initial/target pair (right). \label{fig:C60sym}}
\end{figure}

We considered transport by quantum walk across \C60 with superposition initial and target states chosen to increase the symmetry between the initial and target nodes.  Figure \ref{fig:C60} shows two possible orientations with the initial and target nodes forming a face of the structure, as marked in blue.
The transport from a single node to the diametrically opposite node shown in figures \ref{fig:c60levels1} and \ref{fig:qwC60allcoins} has less symmetry than starting on a pentagonal face or a hexagonal face.  It also has ten levels, rather than eight for the face-terminated orientations.  A proper comparison of transport on these configurations must make allowance for this.  We could give the ten-level systems a two-level head start.  We can also compare with the corresponding cycles, $C_{14}$ and $C_{18}$, to provide a benchmark for the performance.  Both give the same result, so we present the comparison with the cycles here.  We find that the more symmetric initial states are more efficient, with the more symmetric hexagonal face slightly better than the pentagonal face.  The details are shown in figure \ref{fig:qwC60}, to be compared with figure \ref{fig:qwC60allcoins}, using the line for the cycles $C_{14}$ and $C_{18}$ respectively for calibration. 
Figure \ref{fig:C60sym} shows the structure end on, illustrating how the hexagonal face has higher symmetry compared with the pentagonal faces, which are inverted with respect to each other.

\subsection{Transport on nanotube structures}
We now turn to our studies of quantum walk transport on graphene structures.  There are two distinct ways to join up a sheet of graphene to form a tube, depending on whether the ``zig-zag'' pattern runs round the tube or lengthways along the tube.  The pattern orthogonal to the zig-zag direction is known as ``arm-chair''.  Thus we have zig-zag nanotubes with the zig-zag running around the tube, and arm-chair nanotubes, with the zig-zag running along the length of the tube.  Both of these were tested and compared.  There is a third way to join up graphene into nanotubes, where the zig-zag runs obliquely.  We did not test nanotubes of this type in this study, because there are a rather large number of possibilities, all with less symmetry than the chosen configurations, and we have already demonstrated that symmetry significantly enhances transport on \C60.

\begin{figure}
	\begin{minipage}{\columnwidth}
		(a)
		\includegraphics[width=0.7\columnwidth]{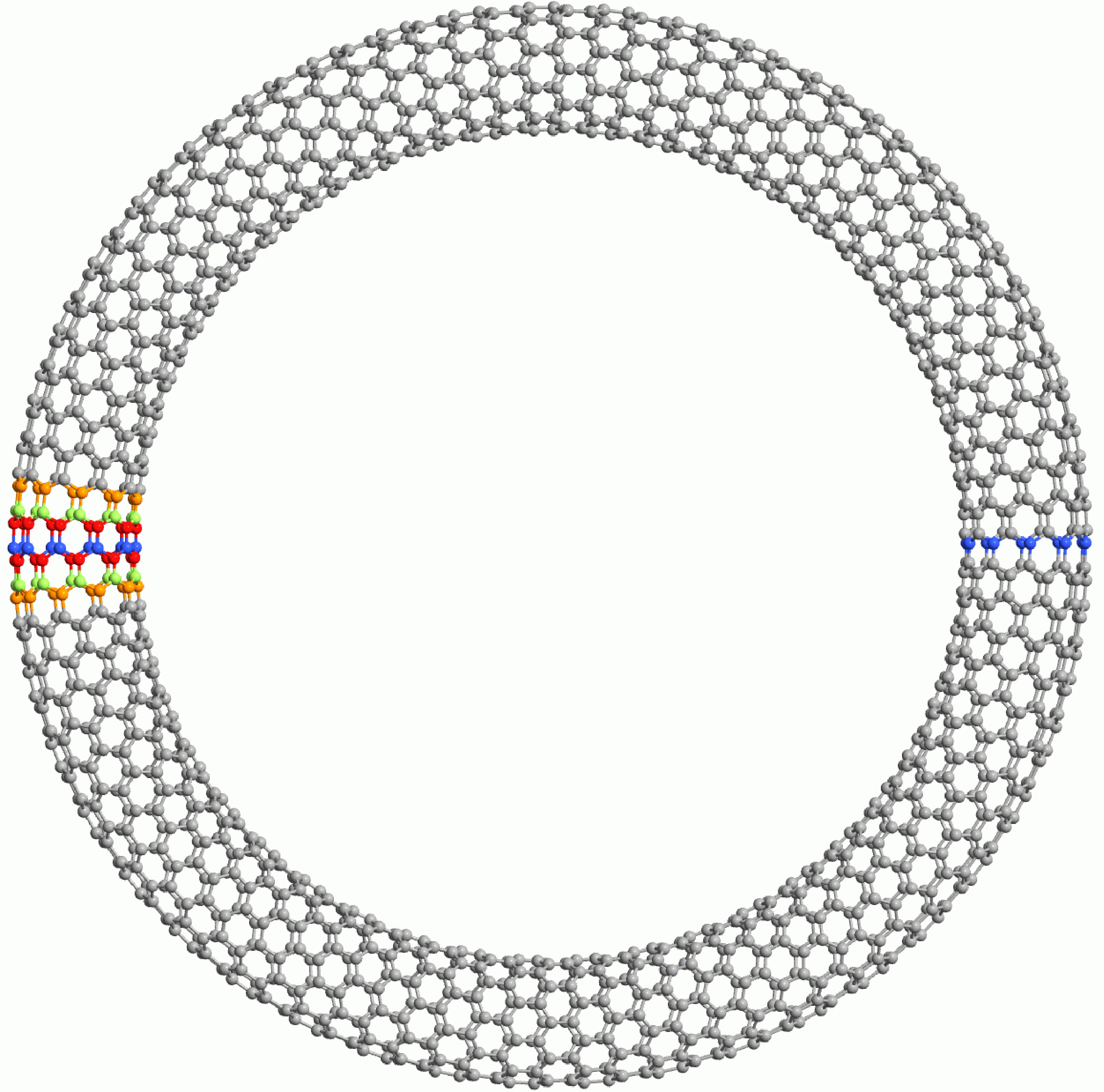}
	\end{minipage}\\
	\begin{minipage}{\columnwidth}
		(b)
		\includegraphics[width=0.7\columnwidth]{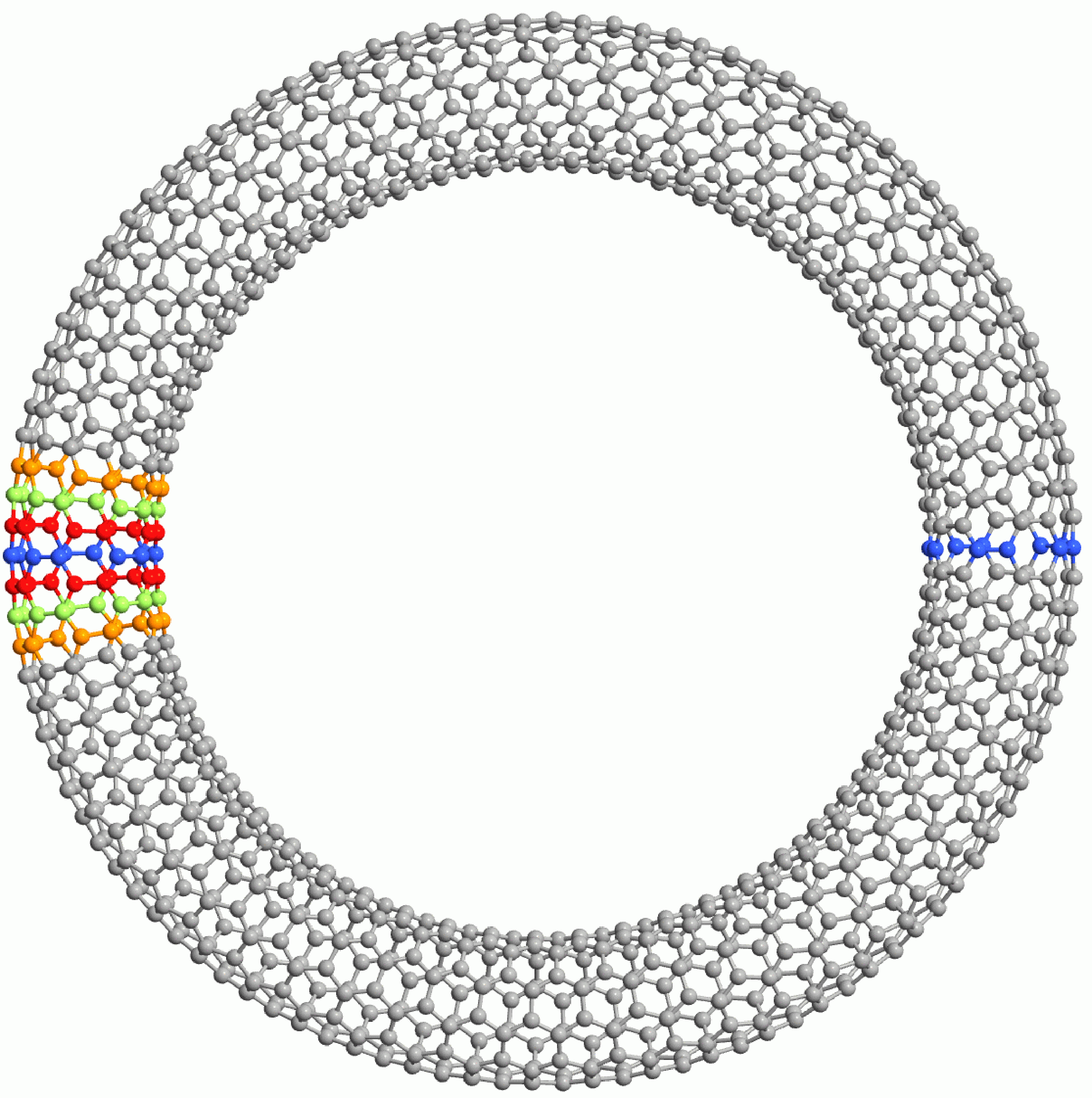}
	\end{minipage}\\
	\caption{(Color online.) (a) zig-zag carbon nanotube loop with 90 repeats forming 91 levels, and (b) arm-chair loop with 55 repeats forming 56 levels, from the sets of initial and target nodes marked in blue A few of the nodes are colored (shaded in print) to indicate the mapping to levels, compare figure \ref{fig:cycle}. \label{fig:nanotube}}
\end{figure}

In order to compare with the cycle, we joined the ends of the nanotubes to form a torus, and studied the transport from one position on the ring to the opposite side, see figure \ref{fig:nanotube}.
The distance from the start is projected onto a line segment in the same way as for the cycle shown in figure \ref{fig:cycle}, so we can track the progress from the initial nodes to the target nodes.  

\begin{figure}
	\includegraphics[width=1.0\columnwidth]{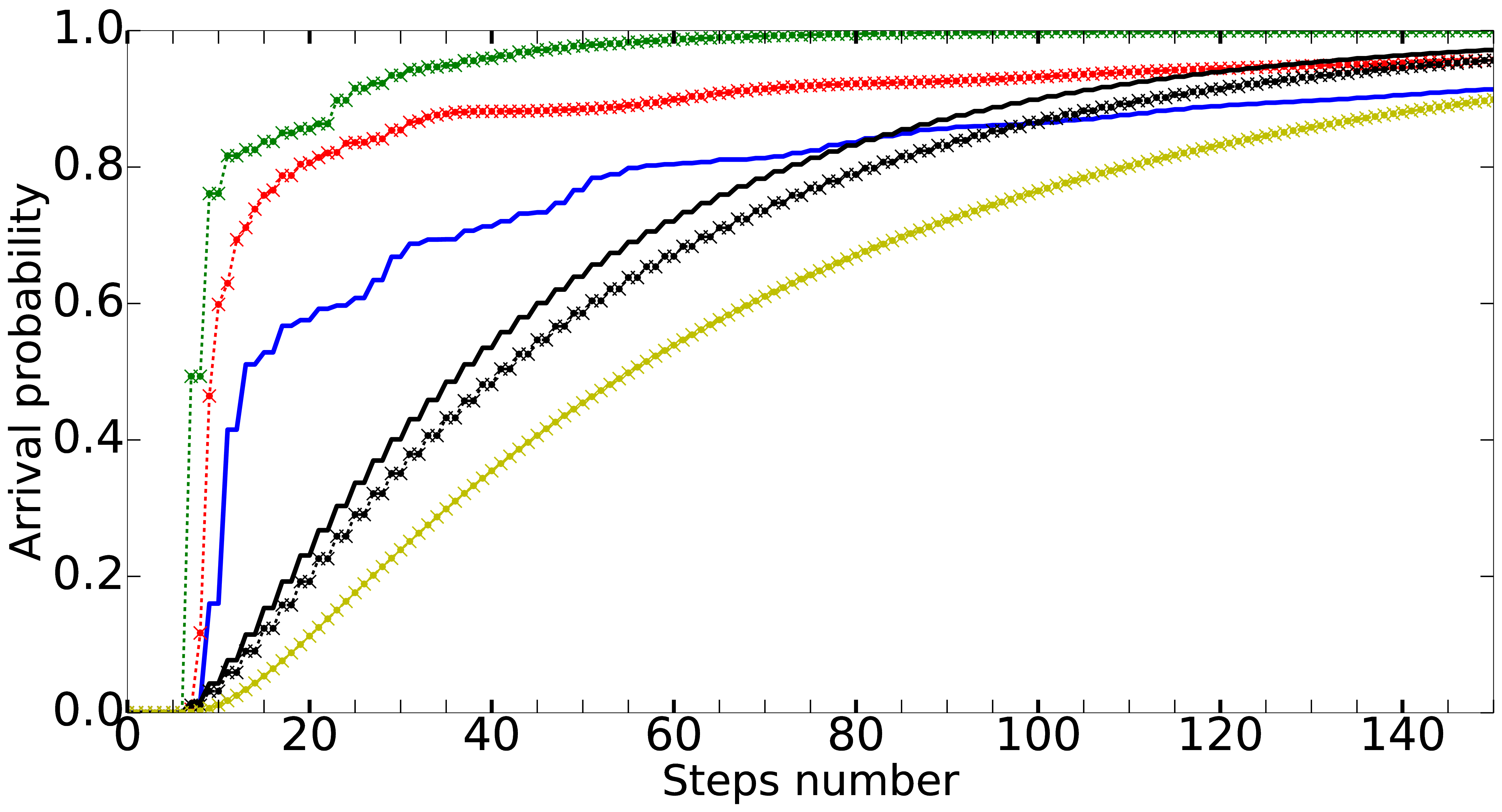}
	\caption{(Color online.)  A quantum walk of 150 steps on loops of zig-zag (green, darker mid grey in print) and arm-chair (red, lighter mid grey in print) carbon nanotube with diameters of 6 ($\times$), 10 (dashes) and 14 (circles), and length corresponding to 8 levels. Cycle $C_{14}$ (blue, dark grey in print) shown for comparison.  Corresponding classical random walks shown in black (zig-zag) and yellow (arm-chair, light grey in print), classical random walk on $C_{14}$ (black, solid).\label{fig:8nano}}
\end{figure}

The results for various diameters of zig-zag and arm-chair nanotube are compared in figure \ref{fig:8nano}, with the results for a cycle $C_{14}$ to provide a benchmark comparison.
The first thing to note is that the diameter of the nanotubes does not affect the transport properties in this setting.  Nanotubes of diameter six, ten and fourteen all gave identical results for the same length.  Next we note that both forms of nanotube are consistently better at transporting the quantum walk to the target nodes than the cycle $C_{14}$.  The zig-zag nanotube shows faster transport than the arm-chair nanotube, with both approaching unit probability eventually, see appendix. 

\begin{figure}
	\includegraphics[width=1.0\columnwidth]{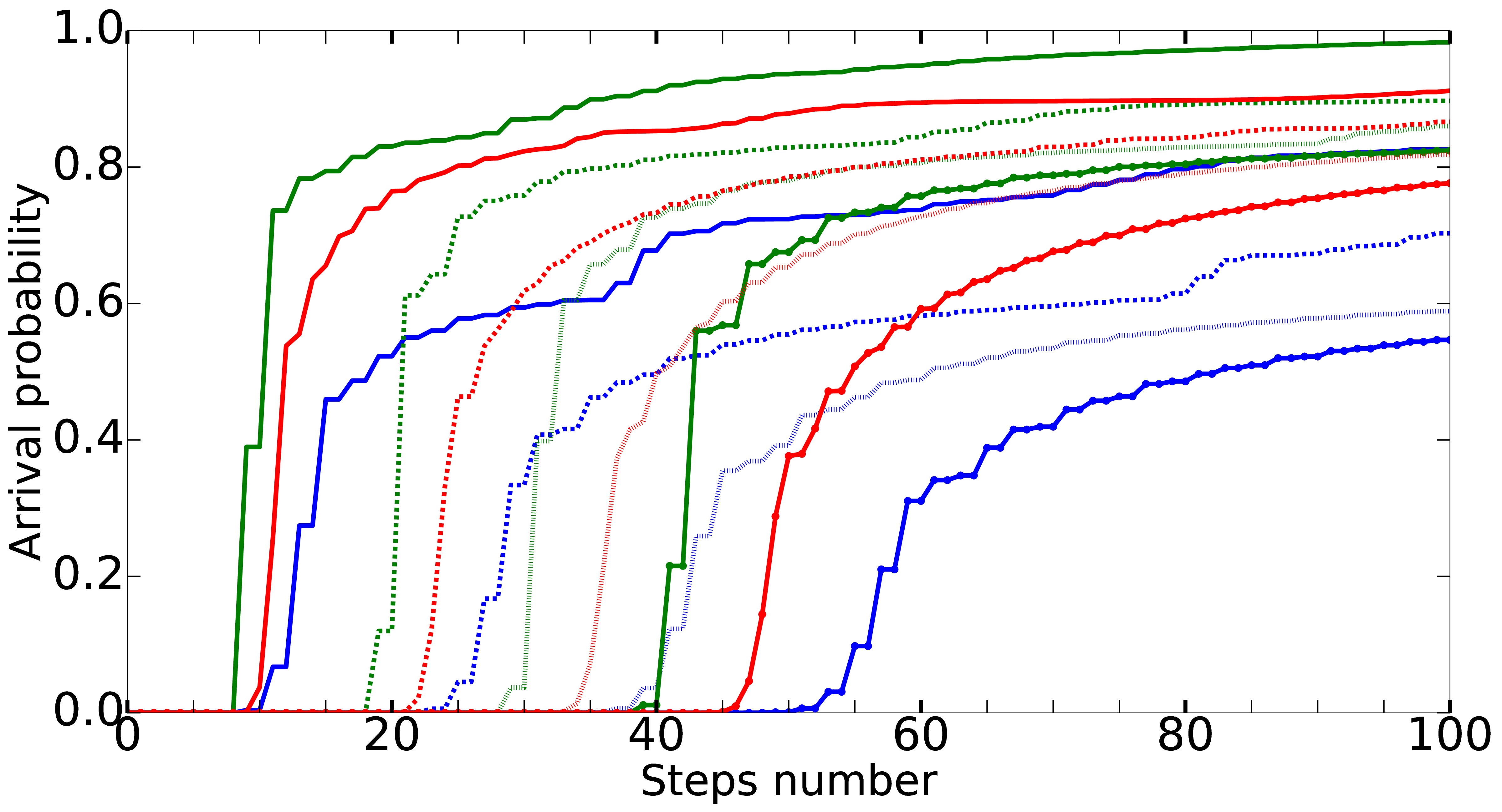}
	\caption{(Color online.) A quantum walk on loops of carbon nanotube of length 10 (solid), 20 (dashed), 30 (dotted), 40 (circles, solid) levels for zig-zag (green, darker mid grey in print) and arm-chair (red, lighter mid grey in print).  Cycles (blue, dark grey in print) of corresponding lengths shown for comparison.\label{fig:lenano}}
\end{figure}

To confirm the supremacy of the arm-chair nanotubes more generally, we also tested different lengths of nanotube, see figure \ref{fig:lenano}.
The pairs of zig-zag (green) and armchair (red) plots can be seen to rise further apart for larger loops, indicating that the zig-zag nanotube loops are providing faster transport over the equivalent lengths.  Both consistently out-perform the cycles of equivalent lengths.

\subsection{Transport on capped nanotubes}
\begin{figure}
	\includegraphics[width=1.0\columnwidth]{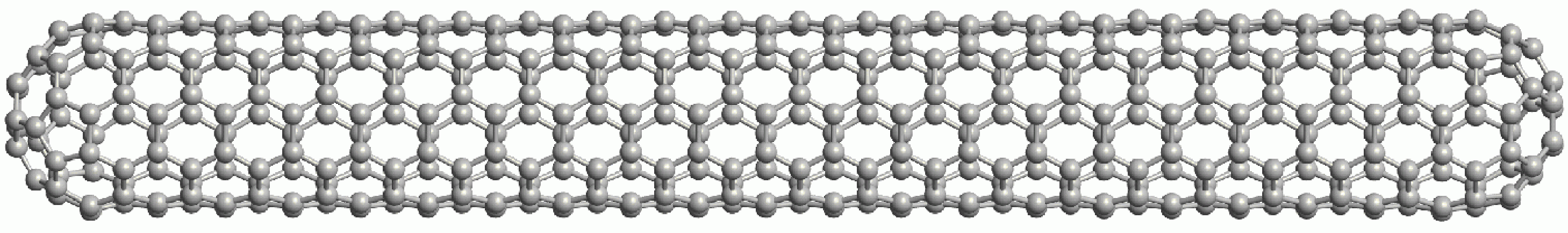}	
	\caption{An armchair nanotube with caps, a pentagonal face at each apex. For a zig-zag nanotube the apices are hexagons (not shown).  The level structure is as for the nanotube loops, but with only one round of carbon atoms per level. \label{fig:capentagon}}
\end{figure}

Joining the nanotubes into loops like cycles puts a strain on the nanotubes and is not a natural form in which they occur.  Bare ends of nanotubes can be irregular, which will not help with efficient coupling or transport.  For nanotubes of matching diameter, a cap of half a \C60 structure can be attached to the ends.  This is like having an elongated \C60 molecule, and the zig-zag or arm-chair character determines whether the end of the cap is a pentagon or hexagon.  An example with an arm-chair configuration that has pentagons at the ends is shown in figure \ref{fig:capentagon}.  

\begin{figure}
	\includegraphics[width=1.0\columnwidth]{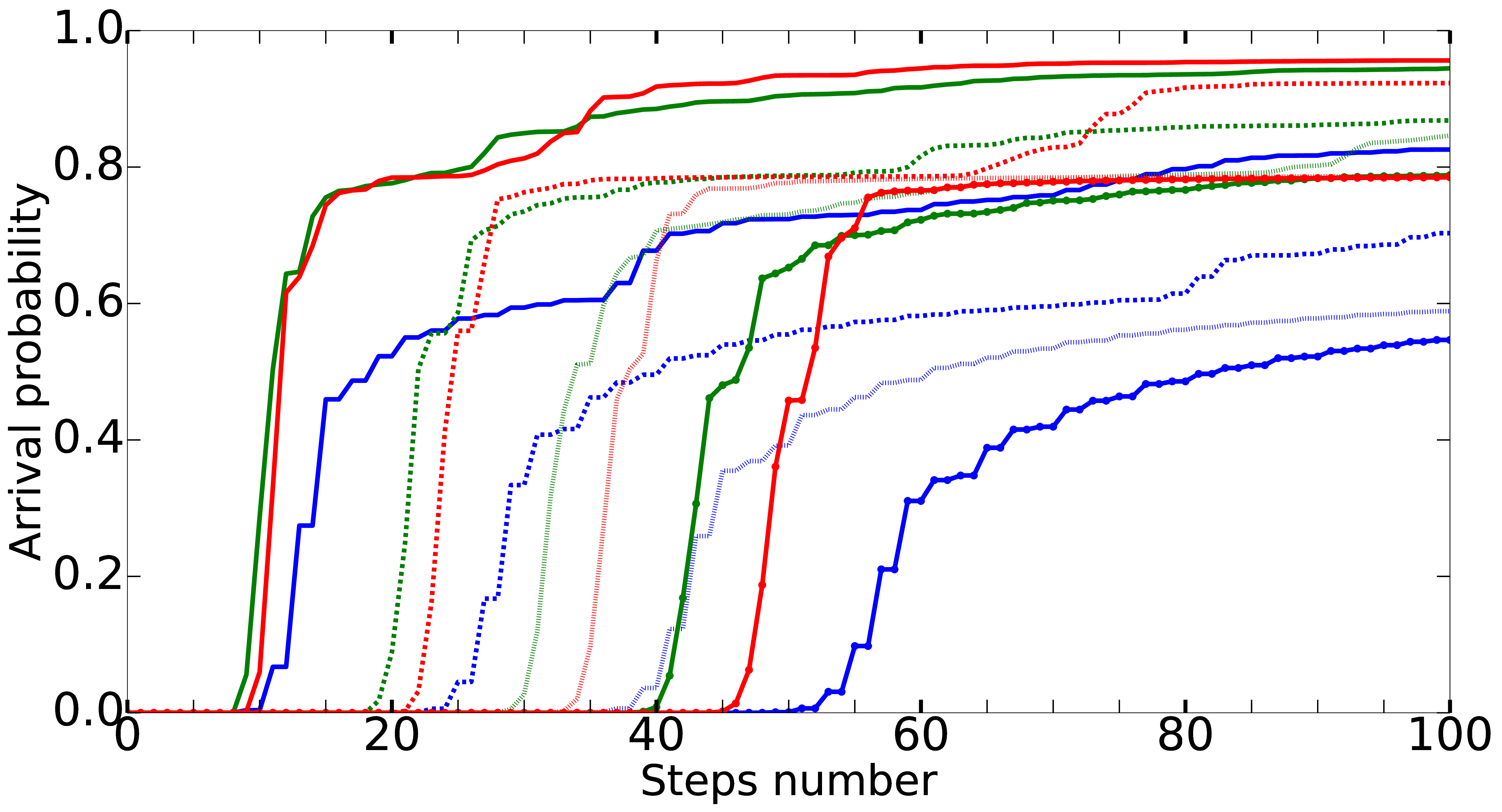}
	\caption{(Color online.) A quantum walk on lengths of capped carbon nanotube: zig-zag (green, darker mid grey in print), armchair (red, lighter mid grey in print) of lengths 10 (solid), 20 (dashed), 30 (dotted), 40 (circles, solid).  Cycles of corresponding lengths shown (blue, dark grey  in print) shown for comparison.\label{fig:capnano}}
\end{figure}

Results for quantum walks on these structures are shown in figure \ref{fig:capnano}.
The zig-zag nanotubes give the fastest transport over short times, as can be seen more clearly for the longer lengths of nanotube.  The arm-chair nanotubes providing the highest arrival probability at later times (the longer lengths were run for more steps to confirm this, not shown in figure \ref{fig:capentagon}).  Both types of nanotubes outperform cycles of the same number of levels.  
\begin{figure}
	\includegraphics[width=1.0\columnwidth]{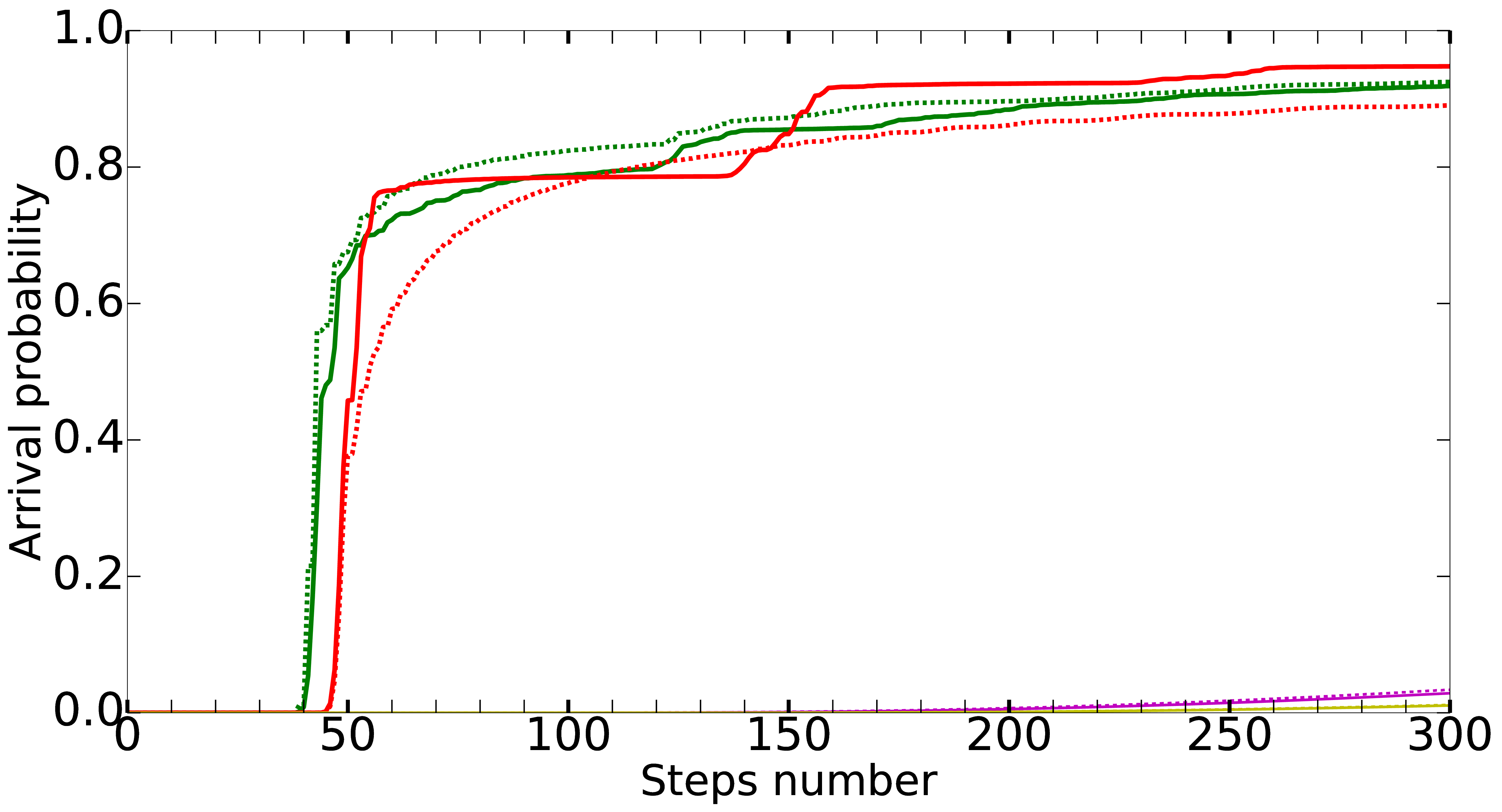}
	\caption{(Color online.) A quantum walk of 300 steps on capped carbon nanotubes zig-zag (green, solid), armchair (red, solid), compared with nanotube loops (dashed).  \label{fig:cap-loop}}
\end{figure}
A comparison between capped nanotubes and nanotube loops is shown in figure \ref{fig:cap-loop}, revealling that the nanotube loops with the same number of levels are slightly better for short lengths, but worse for longer lengths, when the arm-chair capped nanotube approaches unit arrival probability fastest.  This is likely due to the contribution from the caps reducing as a proportion of the total length for longer nanotubes.  The long time behaviour of both is shown in the appendix.
\subsection{Scaling of Transport\label{ssec:scaling}}

Let us now consider the scaling of the transport rate on loops and capped nanotubes. To do this analysis, we examine the number of steps when the probability of arrival exceeds $50\%$, $N_{0.5}$ as a function of the number of levels in the structure. This captures short time behaviour where figures 11 to 14 indicate that simple scaling might be obtained. This analysis allows us to differentiate between faster rates of transport on different structures (appearing as the slope of this quantity) and constant shifts in this quantity which may be caused by the formation of the wavefronts which propagate the walker. In addition to providing qualitative information, we perform linear fits to this data which provide quantitative measurements of the transport rates.
\begin{figure}
	\includegraphics[width=1.0\columnwidth]{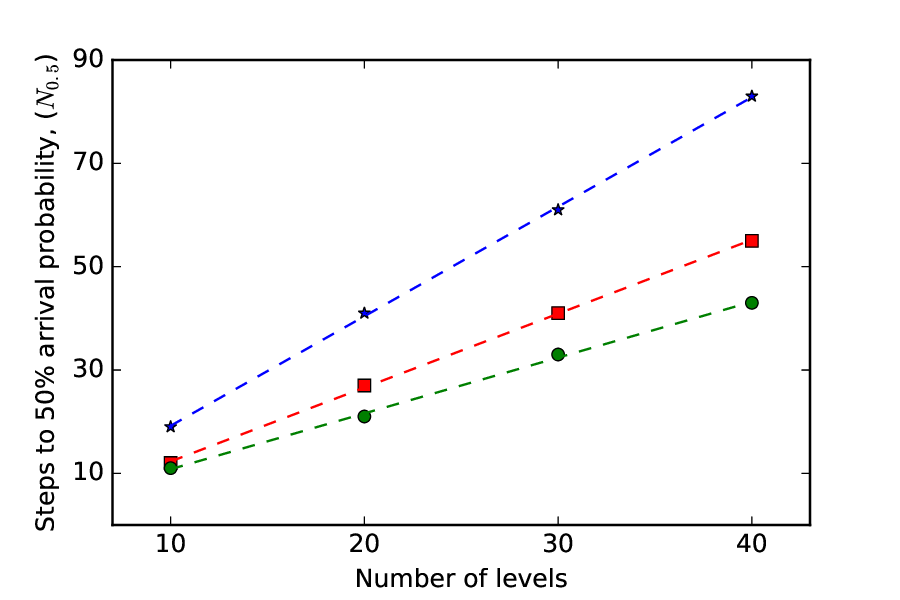}
	\caption{(Color online.)  Number of steps until $50\%$ arrival probability versus number of levels for zig-zag (green circles) and armchair (red squares) loop structures and  the cycle (blue stars). Dashed lines are linear fits, the numbers extracted from these fits are summarized in table \ref{tab:fit_values}. \label{fig:loop_scaling}}
\end{figure}

Let us first consider the nanotube loops, for which the number of steps required for a $50\%$ probability of arrival $N_{0.5}$ is plotted in figure \ref{fig:loop_scaling}. As we can see from this scales linearly with the number of levels, but does so at a different rate for different structures, therefore the difference between armchain, ziz-zag and loop geometries grows linearly with the number of levels. In fact by this metric, transport on a ziz-zag nanotube loop is almost twice as fast as  on the cycle (see table \ref{tab:fit_values} for numbers extracted by numerical fitting). 

\begin{figure}
	\includegraphics[width=1.0\columnwidth]{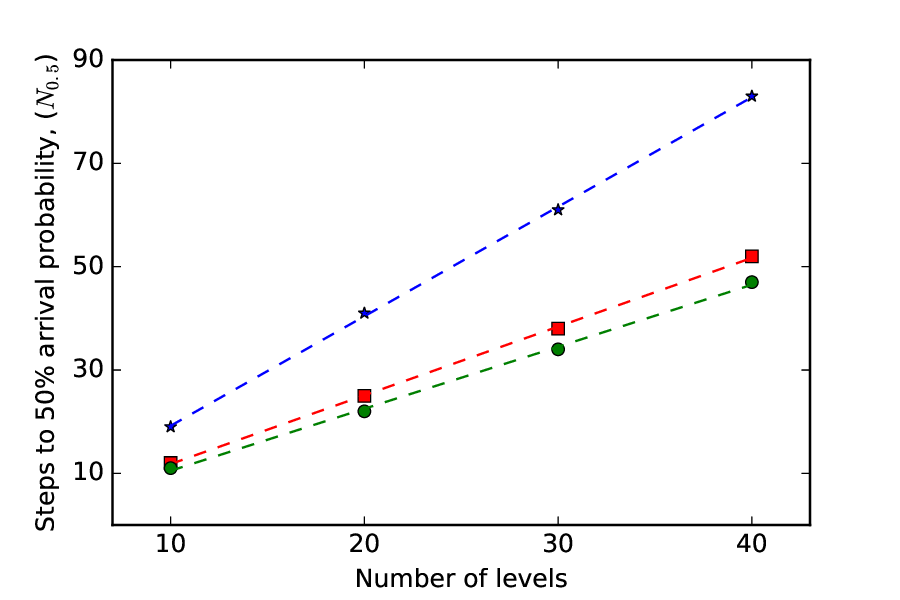}
	\caption{(Color online.)  Number of steps until $50\%$ arrival probability versus number of levels for zig-zag (green circles) and armchair (red squares) capped nanotube structures and  the cycle (blue stars). Dashed lines are linear fits, the numbers extracted from these fits are summarized in table \ref{tab:fit_values}. \label{fig:capped_scaling}}
\end{figure}

We find similar results for the capped nanotubes, as depicted in figure \ref{fig:capped_scaling}, only we find that the difference between zig-zag and armchair is less dramatic. The data in table  \ref{tab:fit_values} reveal that this is due to a combination of the fact that considering a capped rather than loop geometry makes transport on zig-zag nanostructures slower, but makes the transport faster on armchair structures.

\begin{table}
\begin{tabular}{|c|c|c|c|}
\hline 
Structure & m & b & $r^{2}$\tabularnewline
\hline 
\hline 
cycle & 2.12 & -2.00 & 0.9996\tabularnewline
\hline 
loop: ziz-zag & 1.08 & 0.00 & 0.9986\tabularnewline
\hline 
loop: armchair & 1.43 & -2.00 & 0.9997\tabularnewline
\hline 
capped: zig-zag & 1.20 & -1.50 & 0.9986\tabularnewline
\hline 
capped: armchair & 1.33 & -1.50 & 0.9996\tabularnewline
\hline 
\end{tabular}

\caption{Values extracted from linear fit $y=m\,x+b$ for the data in figures \ref{fig:loop_scaling} and \ref{fig:capped_scaling} and coefficient of determination $r^2\equiv1-\frac{\sum_i(y_i-f_i)^2}{\sum_i(y_i-\bar{y})^2}$, where $f_i$ are the data and $y$ is the fitting function,  for each of the fits  \label{tab:fit_values}}
\end{table}

\section{Summary\label{sec:conc}}

We have demonstrated that transport for discrete time quantum random walks is significantly faster on graphs corresponding to a variety of real world nanostructures than it is on simple cycles. In particular, we have demonstrated this for a \C60 fullerene graph and a variety of configurations of graphene nanotube structures, including nanotubes with fullerine caps. Our results consistently show that the nanostructures provide faster transport. Moreover, we demonstrate that in most cases the walker eventually reaches the marked site with unit probability, thus showing that under many, but not all circumstances these structures do not have the problem of infinite hitting times.

Transport across cycles is faster than on the line \cite{kay10a}, hence we have also demonstrated that these structures provide faster transport than walks on a line. The fact that the coined discrete time quantum walk model exhibits faster transport than a simple line may be a discrete time counterpart of the continuous time effects which allow for ballistic transport on real world carbon nanotubes \cite{White98a,Mintmire92a,Hamada92a,Saito92a}. 

The behaviour on these materials can be traced back to the massless behaviour of electrons at the Dirac point in the graphene bandstructure \cite{Novoselov05a}. On the other hand, a discrete time random walk does not carry with it an inherent notion of momentum or energy, so a band structure cannot be defined.  In future work it would be interesting to examine the possibility of a connection between the transport behaviour we see and the bandstructure of these materials.

We have also compared transport behaviour between different nanostructures and have found that zig-zag nanotubes exhibit faster transport than their armchair counterparts. In contrast, theoretical results for electrons in carbon nanotubes show that zig-zag nanotubes behave as semiconductors, while armchair nanotubes behave as metals \cite{Hamada92a,Saito92a}. Therefore in this respect, the  relative transport efficiency between these structures for discrete time random walks are qualitatively different from what is seen in electron transport. We further see that for capped nanotube structures the arrival probability for a zig-zag nanotube always approaches unity, while it does not for the armchair, again indicating better transport for the zig-zag structures. 

It would be interesting in future work to examine which exhibits faster transport for continuous time quantum walks with the same starting conditions we use. On one hand, such walks would be subject to the bandstructure of the continuous material, but would be strongly out of equilibrium, unlike the cases typically examined in electronic transport calculations. It would further be interesting to perform the same calculations as here, but with chiral nanotubes structures and compare with electronic behaviours to investigate further the relationship between electronic and quantum walk properties.

\begin{acknowledgments}	
HB supported by the scholarship program of the Algerian Ministry of Higher Education and Scientific Research (459/PNE/ENS/GB/2015-2016), and Tebessa University.  NC and VK supported by the UK Engineering and Physical Sciences Research Council Grant EP/L022303/1.
\end{acknowledgments}
%
\appendix*
\section{Long time behaviour of quantum walks}
%
\begin{figure}[!htb]
	\includegraphics[width=1.0\columnwidth]{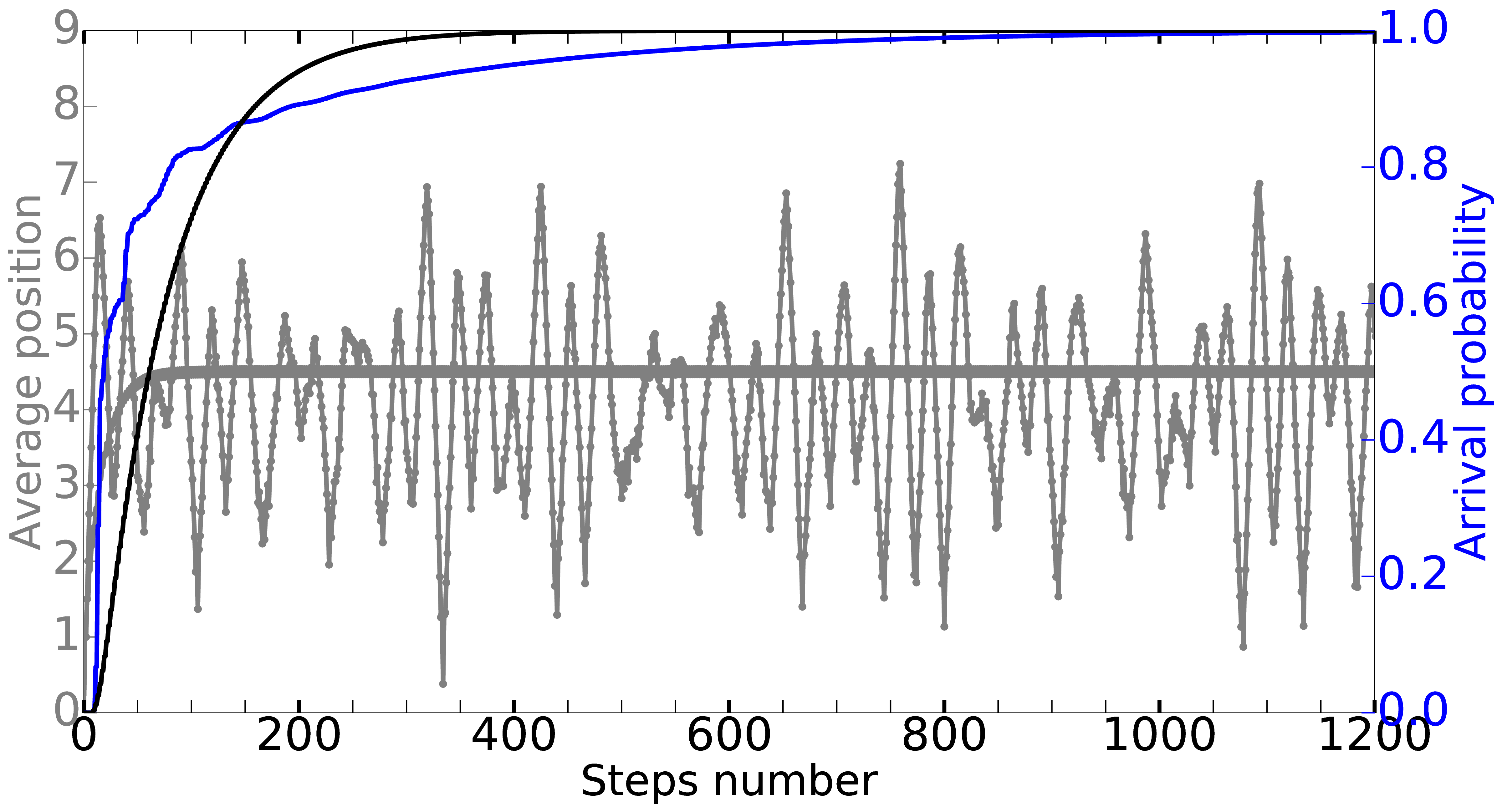}
	\caption{(Color online.) Quantum and classical random walks on $C_{18}$ for 1,200 steps showing arrivial times converge to unity for both, while the oscillations about 4.5 for the average position continue indefinitely for the quantum walk.  Compare figure \ref{fig:qwcycle}\label{fig:C18-1.2K}}
\end{figure}
The long term behaviour of the quantum walks studied in this paper is presented here, in particular, whether the probability to arrive at the marked state approaches unity, as far as we can determine this from numerical studies. As figure \ref{fig:C18-1.2K} shows, this always happens for the cycle without a `wait' state. Furthermore, if we don't measure, we see that the classical walk converges to a constant probability to be found on the marked site, while the quantum probability continues to fluctuate for all time.

\begin{figure}[!htb]
	\includegraphics[width=1.0\columnwidth]{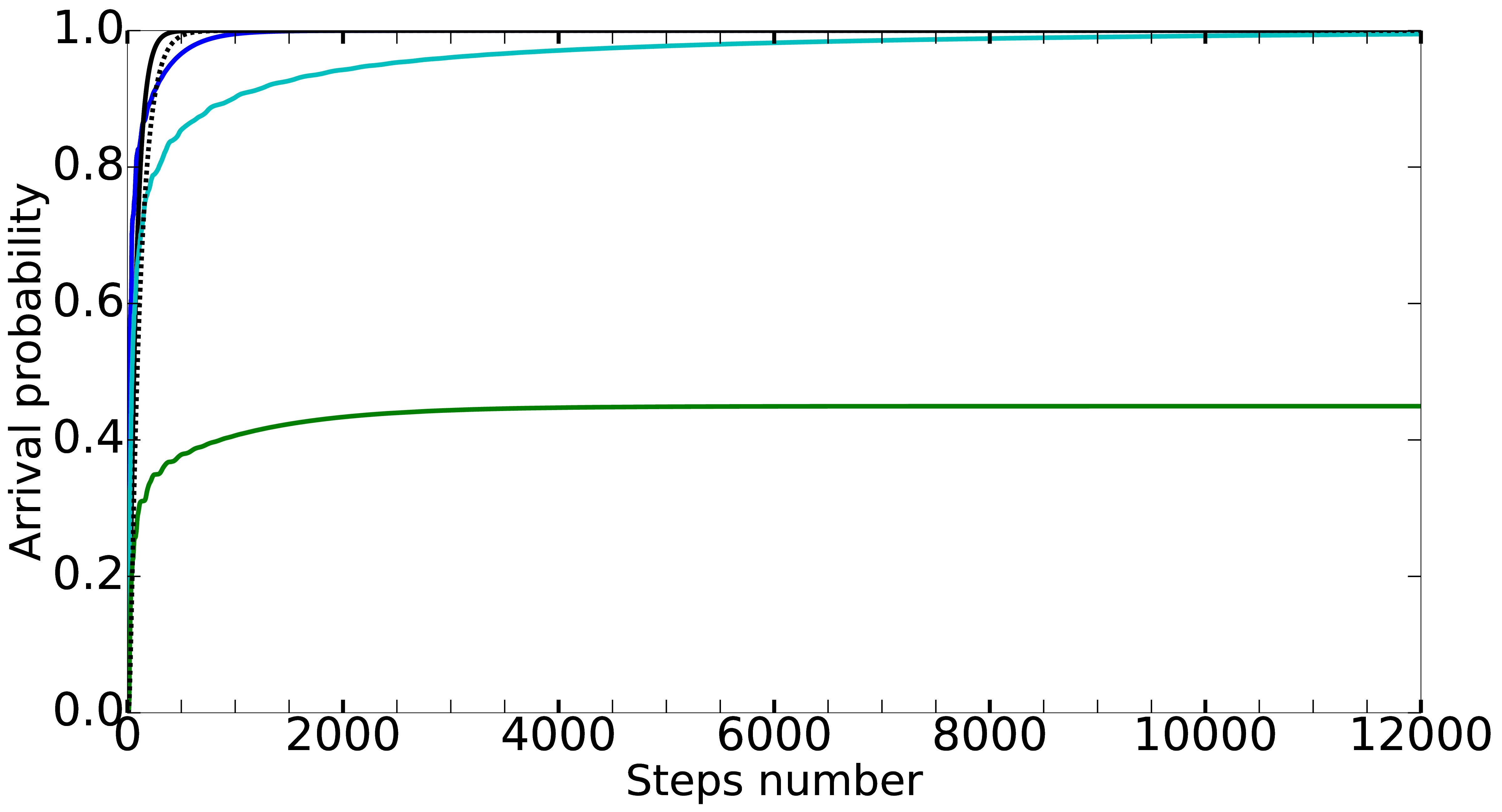}
	\caption{(Color online.) Quantum and classical random walks on $C_{18}$ for 5,000 steps showing the 2D coins converge to unity, while the 3D coins do not in the quantum case, but do in the classical case. 5,000 steps of a quantum walk on a cycle $C_{18}$ with $18$ nodes mapped to ten levels from start ($0$) to target node ($9$), see figure \ref{fig:cycle}.  Arrival probability plotted against the number of time steps for coin operators $H$ (blue), $G_3$ (green), $F_3$ (turquoise).  A classical random walk using an unbiased 2-sided (black, solid) and 3-sided coin (black, dotted) shown for comparison.\label{fig:C18-12K}}
\end{figure}
\begin{figure}[!htb]
	\includegraphics[width=1.0\columnwidth]{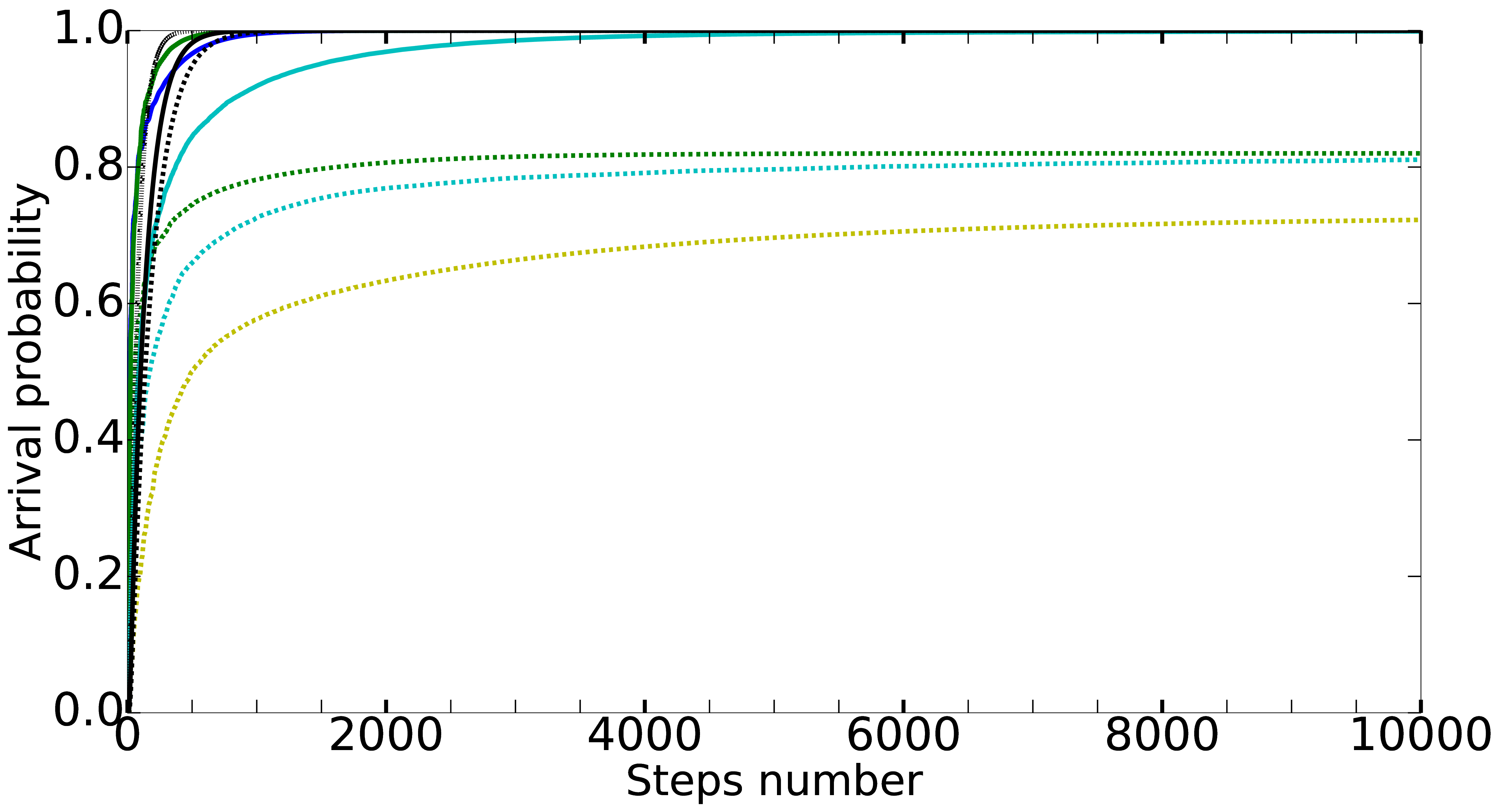}
	\caption{(Color online.) A quantum walk of 10,000 steps on \protect{\C60} between single opposite nodes, mapped to ten positions as in figure \ref{fig:c60levels1}.  Arrival probability plotted against the number of time steps for coin operators $G_3$ (green, solid, dark grey in print), $G_4$ (green, dashed, mid grey in print), $F_3$ (turquoise, solid, light grey in print), $F_4$ (turquoise, dashed, light grey in print), $H\otimes H$ (yellow, dashed, very light grey in print).  A cycle $C_{18}$ is shown (blue, dark grey in print) along with classical random walks for a 3-sided (black, solid) and 4-sided (black, dashed) coin and the cycle (black, dotted) are shown for comparison.\label{fig:qwC60allcoinslong}}
\end{figure}
\begin{figure}[!htb]
	\includegraphics[width=1.0\columnwidth]{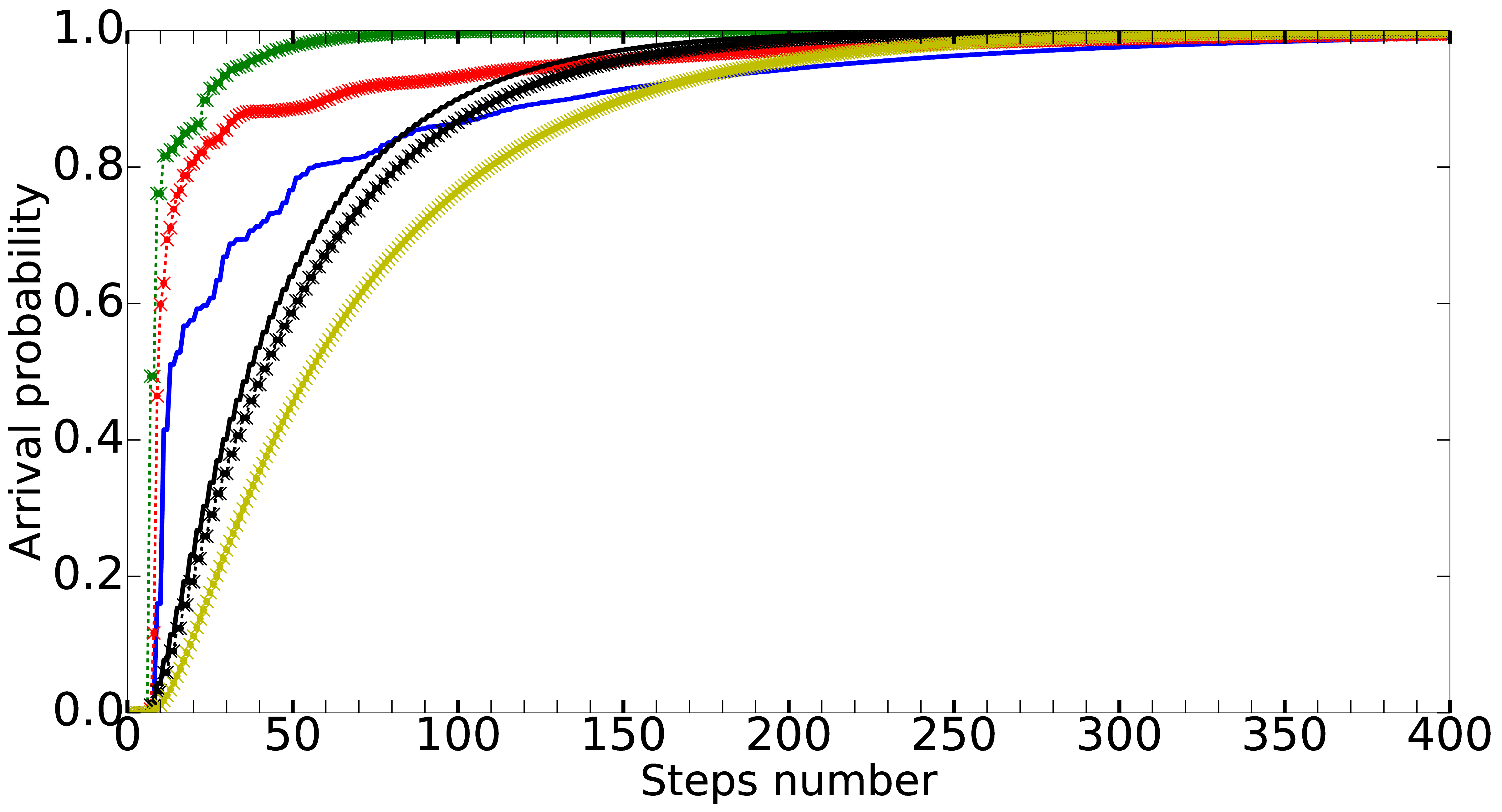}
	\caption{(Color online.)  A quantum walk of 400 steps on loops of zig-zag (green, darker mid grey in print) and arm-chair (red, lighter mid grey in print) carbon nanotube with diameters of 6 ($\times$), 10 (dashes) and 14 (circles), and length corresponding to 8 levels. Cycle $C_{14}$ (blue, dark grey in print) shown for comparison.  Corresponding classical random walks shown in black (zig-zag) and yellow (arm-chair, very light grey in print), classical random walk on $C_{14}$ (black, solid).\label{fig:8nanolong}}
\end{figure}
\begin{figure}[!htb]
	\includegraphics[width=1.0\columnwidth]{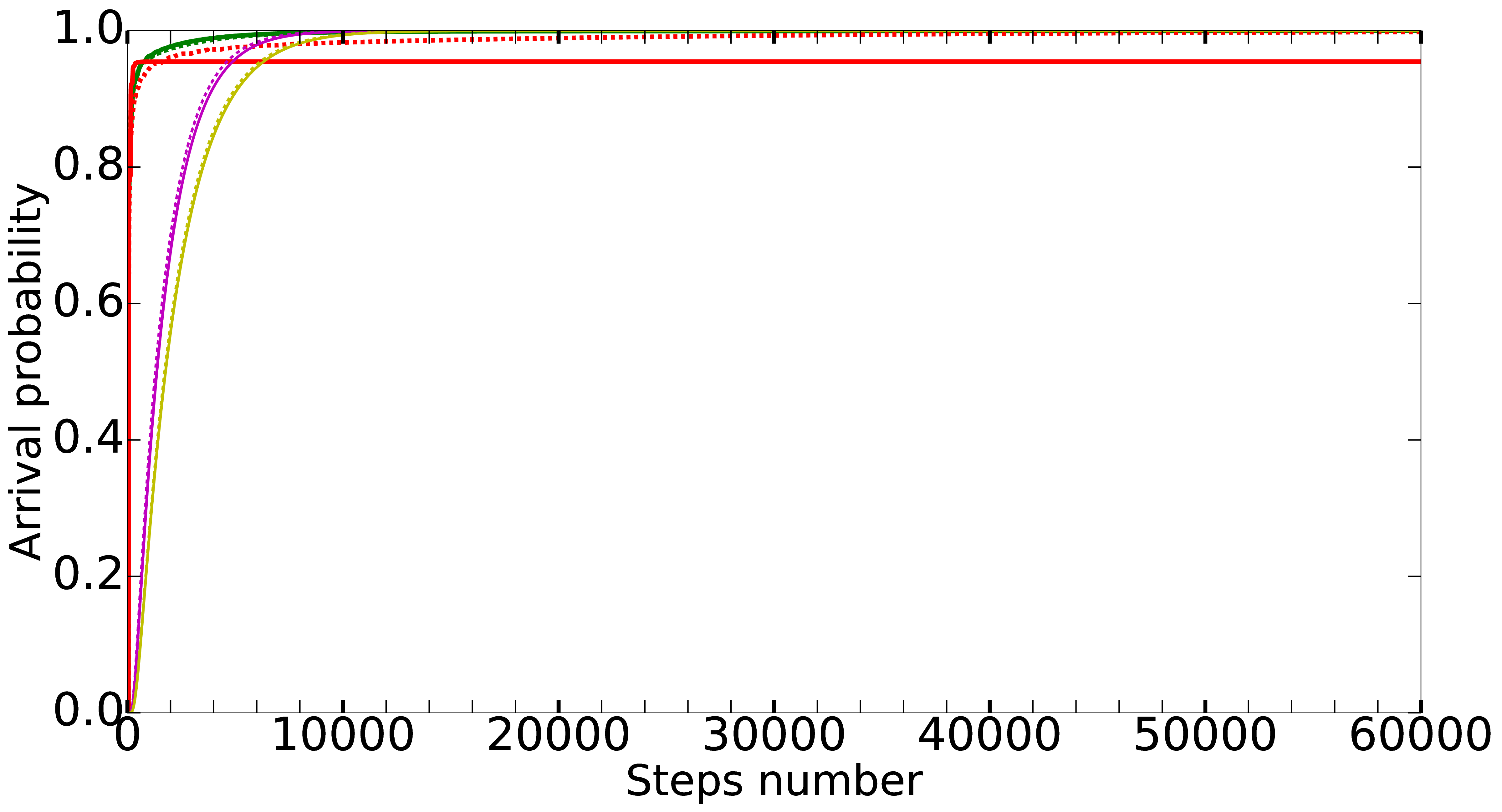}
	\caption{(Color online.) A quantum walk of 60,000 steps on capped carbon nanotubes zig-zag (green, solid, darker mid grey in print), armchair (red, solid, lighter mid grey in print), compared with nanotube loops (dashed), and corresponding classical random walks for zig-zag (pink, dark grey in print) and armchair (yellow, light grey in print), to show the long time behavior.  \label{fig:cap-looplong}}
\end{figure}

On the other hand, figure \ref{fig:C18-12K} shows that on the cycle, convergence to unity does not occur for the $F_3$ coin which includes a wait state, although it does for some others which include a wait state. This parallels some of the behaviour for \C60 starting on a single site, with a single marked site.  As illustrated in figure \ref{fig:qwC60allcoinslong}, the arrival probability converges to unity for coins without wait states, but does not for coins which do have them.
We see similar behaviour for the nanotube loop arrival probability depicted in figure \ref{fig:8nanolong}, which depicts the results for a nanotube loop with using a $G_3$ coin, and for which arrival probabilities all approach unity. However, as figure \ref{fig:cap-looplong} illustrates, this is not true for all cases of capped nanotubes, in particular this probability does not approach unity for the capped armchair nanotube. It is also interesting to note that for the uncapped version of the armchair nanotube, the arrival probability does approach unity, but it does so much slower than either the zig-zag nanotube or the cycle. This very slow arrival warrants future study.
%
\bibliography{qwCnanotube} 
%
\end{document}